\documentclass[10pt,journal,compsoc]{IEEEtran}

\usepackage{cite}
\usepackage{graphicx}
\usepackage{amsmath,amssymb,amsfonts}
\usepackage{algpseudocode} % For \For and other algorithmic commands
\usepackage{algorithm}
\usepackage{textcomp}
\usepackage{xcolor}
\usepackage{multirow}
\usepackage{booktabs}
\usepackage{url}
\usepackage{hyperref}
\hypersetup{colorlinks=true,linkcolor=blue,citecolor=blue,urlcolor=blue}

\usepackage{stfloats}
\usepackage{url}
\usepackage{verbatim}
\usepackage{threeparttable}
\usepackage{tabularx}
\usepackage{longtable}
\usepackage{graphicx}
\usepackage{multirow}
\usepackage{pifont}
\usepackage{titlesec}
\usepackage{booktabs}
\usepackage{algpseudocode}
\usepackage{graphicx}
\usepackage{subcaption}
\usepackage{enumitem}
\usepackage{amsmath}
\usepackage{enumitem}
\usepackage{ragged2e}

\begin{document}

\title{Towards Reliable Service Provisioning for Dynamic UAV Clusters in Low-Altitude Economy Networks}

\author{Yanwei~Gong, Ruichen~Zhang, Xiaoqing Wang, Xiaolin Chang, Bo Ai,~\IEEEmembership{Fellow,~IEEE},  Junchao Fan, Bocheng Ju, and Dusit~Niyato,~\IEEEmembership{Fellow,~IEEE}
\IEEEcompsocitemizethanks{
\IEEEcompsocthanksitem Yanwei Gong, Xiaoqing Wang, Xiaolin Chang, Junchao Fan, Bocheng Ju are with the Beijing Key Laboratory of Security and Privacy in Intelligent Transportation, Beijing Jiaotong University, P.R.China. (e-mail: xlchang@bjtu.edu.cn)
\IEEEcompsocthanksitem Bo Ai is with the School of Electronics and Infomation Engineering, Beijing Jiaotong University, P.R.China. (e-mail: boai@bjtu.edu.cn)
\IEEEcompsocthanksitem Ruichen Zhang and Dusit Niyato are with the College of Computing and Data Science, Nanyang Technological University, Singapore.(e-mail: \{ruichen.zhang, dniyto\}@ntu.edu.sg)
}}

\IEEEtitleabstractindextext{
\begin{abstract}
\justifying
Unmanned Aerial Vehicle (UAV) cluster services are crucial for promoting the low-altitude economy by enabling scalable, flexible, and adaptive aerial networks. To meet diverse service demands, clusters must dynamically incorporate a New UAVs (NUAVs) or an Existing UAV (EUAV). However, achieving sustained service reliability remains challenging due to the need for efficient and scalable NUAV authentication, privacy-preserving cross-cluster authentication for EUAVs, and robust protection of the cluster session key, including both forward and backward secrecy. To address these challenges, we propose a Lightweight and Privacy-Preserving Cluster Authentication and Session Key Update (LP2-CASKU) scheme tailored for dynamic UAV clusters in low-altitude economy networks. LP2-CASKU integrates an efficient batch authentication mechanism that simultaneously authenticates multiple NUAVs with minimal communication overhead. It further introduces a lightweight cross-cluster authentication mechanism that ensures EUAV anonymity and unlinkability. Additionally, a secure session key update mechanism is incorporated to maintain key confidentiality over time, thereby preserving both forward and backward secrecy. We provide a comprehensive security analysis and evaluate LP2-CASKU performance through both theoretical analysis and OMNeT++ simulations. Experimental results demonstrate that, compared to the baseline, LP2-CASKU achieves a latency reduction of \textbf{82.8\%–90.8\%} by across different UAV swarm configurations and network bitrates, demonstrating strong adaptability to dynamic communication environments. Besides, under varying UAV swarm configurations, LP2-CASKU reduces the energy consumption by approximately \textbf{37.6\%–72.6\%}, while effectively supporting privacy-preserving authentication in highly dynamic UAV cluster environments.

\end{abstract}

}

\maketitle

% ================================
% Sections start here
% ================================

\section{Introduction}
% [填入 Introduction 章节内容]
\IEEEPARstart{U}{nmanned} Aerial Vehicles (UAVs), commonly known as drones, have become integral to advancing the low-altitude economy by leveraging near-ground airspace to deliver efficient and real-time services~\cite{liu2025energy, jiang2023ntn}. In particular, UAV clusters enhance task efficiency, coverage scale, and intelligent coordination, supporting a wide range of applications such as aerial photography, precision agriculture, infrastructure inspection, disaster response, and logistics~\cite{ref2}--\cite{zhang2024moe}. Recent advances in technologies such as artificial intelligence, 5G connectivity, and swarm intelligence have further enabled autonomous and large-scale UAV cluster operations, significantly accelerating their integration into smart cities and low-altitude economic infrastructures~\cite{zhang2024generative}--~\cite{zhang2023ppo}.

\begin{figure}[!t]
\includegraphics[width=3.5in]{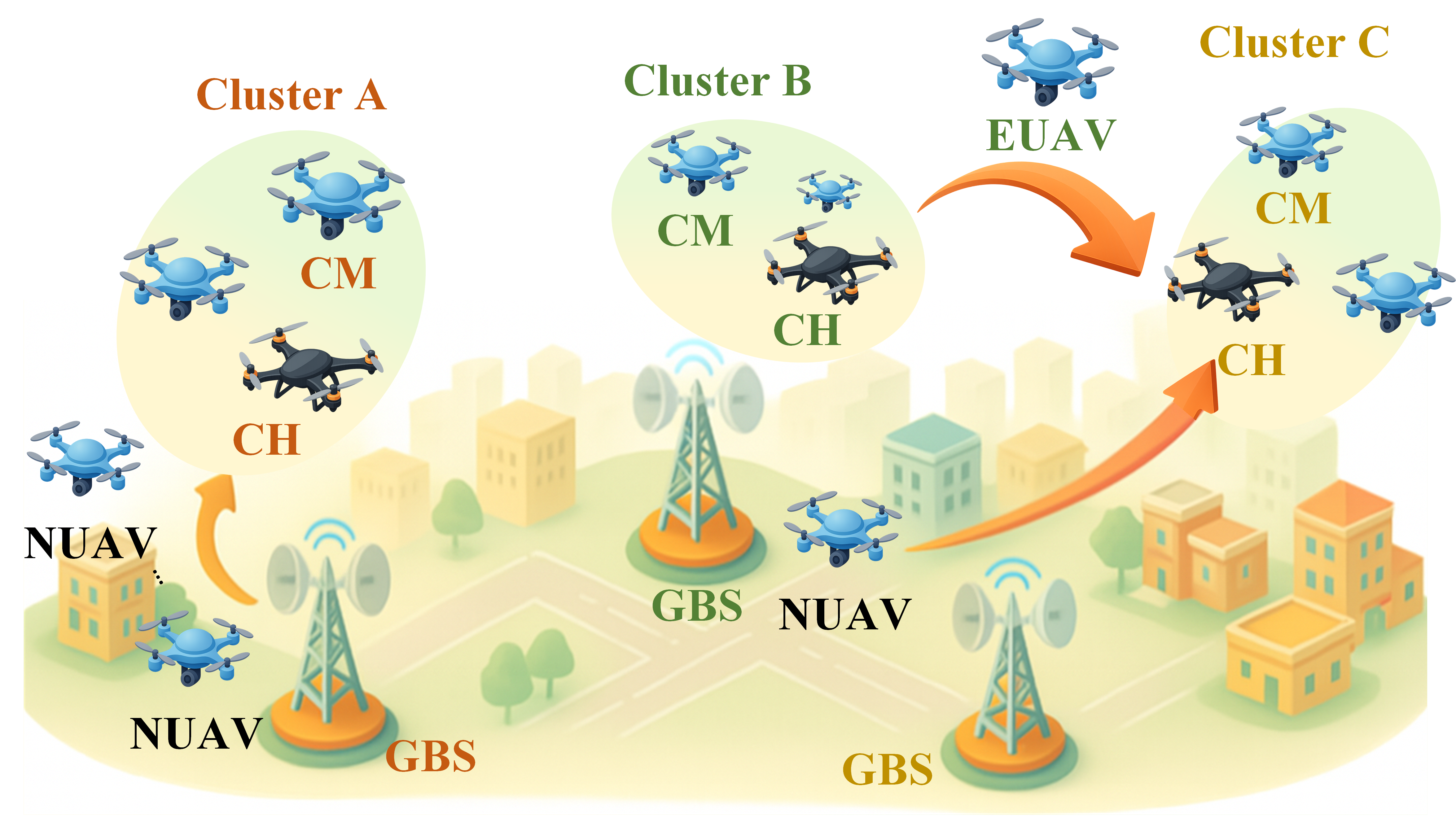}
\caption{Hierarchical UAV swarm architecture for low-altitude economy networking. A UAV swarm consists of multiple UAV clusters, each coordinated by a CH. The architecture enables scalable cluster-based services with dynamic UAV membership management across clusters.}
\label{fig1}
\vspace{-10pt}
\end{figure}

To efficiently deliver these services, UAVs operating in low-altitude economic networks are typically organized into hierarchically structured swarms. As illustrated in Fig.\ref{fig1}\cite{ref8}, each swarm is divided into clusters, each consisting of a Cluster Head (CH) and multiple Cluster Members (CMs). The CH coordinates cluster tasks and manages CMs, leveraging a shared cluster session key to ensure secure internal communications. In addition, Ground Base Stations (GBSs) oversee multiple UAV clusters by distributing instructions to CHs and collecting task results, thereby ensuring system responsiveness, efficiency, and scalability—core requirements for low-altitude economy applications~\cite{ref9}.

Despite these advantages, UAV clusters inherently exhibit high dynamism~\cite{ref10}. For instance, a CM may lose connectivity with its CH or with other CMs, leading to reduced service reliability~\cite{ref11}. Moreover, to meet the real-time, scalable, and dependable service demands of the low-altitude economy, UAV clusters often require dynamic reconfiguration, such as the recruitment of additional UAVs. Typically, a CH may request support from its associated GBS to deploy New UAVs (NUAVs), or coordinate with other clusters to incorporate Existing UAVs (EUAVs). Consequently, ensuring continuous reliability necessitates secure authentication of these UAVs and robust protection of the cluster session key, as the dynamic joining or departure of UAVs may expose the existing session key to potential compromise. These needs introduce several critical challenges, detailed as follows:

\begin{itemize}
    \item \textit{Challenge 1: Efficient and scalable cluster authentication of multiple NUAVs}. To maintain the reliability of a dynamic UAV cluster, the CH and CMs must authenticate each NUAV. Moreover, considering that a NUAV may later join other clusters as a EUAV, the CHs of those clusters must also authenticate it. These processes of authenticating NUAVs is referred to as \textbf{cluster authentication}. However, when multiple NUAVs attempt to join simultaneously, performing individual authentication sequentially incurs substantial communication overhead and latency, as the CH must forward each request to its CMs. This is inadequate for time-sensitive applications in the low-altitude economy. Therefore, developing an efficient batch authentication mechanism that supports simultaneous authentication of multiple NUAVs while minimizing communication costs remains a critical challenge.

    \item \textit{Challenge 2: Efficient cross-cluster authentication with privacy protection}. To maintain service reliability, it is essential to authenticate EUAVs when they join a cluster, a process termed \textbf{cross-cluster authentication}. In addition, since a EUAV may participate in multiple such processes, preserving its privacy is vital to prevent inference attacks on its flight trajectory~\cite{ref12}, a particularly serious concern in urban low-altitude airspace, where both privacy and security risks are critical. To prevent adversaries from linking different authentication sessions and reconstructing the EUAV’s movement patterns, it is necessary to ensure that the authentication messages across clusters cannot be correlated. Achieving this requires that messages exchanged during different cross-cluster authentications of the same EUAV remain unlinkable, thereby preventing the attacker from associating these sessions to a single UAV identity or tracking its path over time. Thus, the challenge is to design an efficient and privacy-preserving cross-cluster authentication mechanism.

    \item \textit{Challenge 3: Forward and backward secrecy of the cluster session key}. Ensuring the secrecy of the cluster session key is critical to the reliability of dynamic UAV cluster services. When NUAVs or EUAVs join a cluster, the cluster session key must be updated to maintain forward secrecy. Similarly, when a CM departs, whether due to failure or is reassigned as a EUAV to another UAV cluster, the cluster session key must be updated to preserve backward secrecy. This ensures that the departing CM can no longer access subsequent intra-cluster communications. These mechanisms are essential to meeting the data security requirements of the low-altitude economy~\cite{cai2025secure}. Therefore, designing a secure session key update mechanism that guarantees both forward and backward secrecy remains a key challenge.
\end{itemize}

Existing authentication and key management schemes for UAV networks~\cite{ref13}--\cite{ref20} face critical challenges in dynamic UAV cluster environments. They often lack efficient support for batch authentication of multiple NUAVs, fail to ensure EUAV privacy through anonymity and unlinkability, and overlook backward secrecy in session key updates. Furthermore, many~\cite{ref13},~\cite{ref14},~\cite{ref17},~\cite{ref18} rely on computationly intensive techniques or incur excessive overhead, making them unsuitable for real-time, resource-constrained low-altitude scenarios. To address these limitations, we propose the first Lightweight and Privacy-Preserving Cluster Authentication and Session Key Update (LP2-CASKU) scheme for dynamic UAV clusters. LP2-CASKU integrates three key mechanisms to enhance authentication efficiency, strengthen privacy protection, and ensure the secrecy of the cluster session key, thereby maintaining reliable UAV cluster services. Its novel features are as follows:

\begin{itemize}
    \item \textbf{Message aggregation for efficient cluster authentication.} LP2-CASKU incorporates a message aggregation mechanism in terms of signature aggregation and public key aggregation, enabling the cluster to authenticate multiple NUAVs simultaneously while reducing communication overhead and latency. This addresses \textit{Challenge 1}.

    \item \textbf{Lightweight cross-cluster authentication with anonymity and unlinkability.} LP2-CASKU introduces a lightweight cross-cluster authentication mechanism that allows the CH to efficiently authenticate EUAVs. It also ensures EUAV anonymity and unlinkability of messages exchanged during different cross-cluster authentications of the same EUAV, thereby preserving privacy and addressing \textit{Challenge 2}.

    \item \textbf{Cluster session key update with forward and backward secrecy.} LP2-CASKU provides a cluster session key update mechanism to ensure the secrecy of the cluster session key. This mechanism guarantees forward secrecy when UAVs join a cluster and backward secrecy when UAVs leave, thereby fulfilling \textit{Challenge 3}.
\end{itemize}

We conduct both formal and illustrative security analyses to validate the proposed LP2-CASKU, and provide theoretical evaluations of its computation and communication overheads. To assess practical feasibility, we conduct OMNeT++~\cite{ref21}-based simulations under realistic UAV configurations by taking multiple realistic factors into consideration, such as UAV flight speed, signal transmission power network communication protocol stack, and so on. Specifically, compared to the baseline, the message aggregation mechanism significantly reduces authentication latency by \textbf{82.8\%–90.8\%} under  varying UAV swarm configurations, and achieves consistent latency reductions of \textbf{88.0\%–89.5\%} across different network bitrates. In parallel, LP2-CASKU reduces energy consumption by \textbf{36.1\%--72.6\%} for CHs and CMs, and by \textbf{40.9\%--70.9\%} for other CHs and NUAVs. These results demonstrate that LP2-CASKU effectively mitigates communication and computation burdens while ensuring efficient, secure, and privacy-preserving authentication in dynamic UAV cluster environments.

The remainder of the paper is organized as follows. Section 2 reviews related works, and Section 3 introduces the preliminaries and system model. Section 4 presents the design of LP2-CASKU. Sections 5 and 6 analyze the security and evaluate the performance, respectively. Section 7 concludes the paper.

\section{Related Work}
% [填入 Related Work 章节内容]
We present related works about UAV authentication in this section. For each work, we discuss its limitations to clarify the motivation of this paper and highlight the advantages of LP2-CASKU. Table~\ref{table1} provides a comparative summary between existing schemes and LP2-CASKU.

Tan et al.~\cite{ref13} proposed an authentication scheme for UAVs using blockchain and smart contracts. However, the blockchain introduced additional latency due to transaction confirmation time, which failed to meet the real-time requirements of low-altitude economy networks. Moreover, cross-cluster authentication was not considered. Feng et al.~\cite{ref14} also designed a blockchain-based authentication scheme for UAVs. Although it supported cross-cluster authentication, it did not include a cluster session key update mechanism. In addition, the latency issue caused by blockchain remained unresolved. Zhang et al.~\cite{ref15} proposed a cluster session key agreement scheme for UAVs. While it satisfied forward and backward secrecy through key updates, it lacked support for cross-cluster authentication. Karmakar et al.~\cite{ref16} introduced a distributed UAV authentication scheme based on blockchain. Their method supported cross-cluster authentication and maintained forward secrecy for session keys, but neglected backward secrecy. Xie et al.~\cite{ref17} designed an authentication protocol for UAV-assisted Internet of Vehicles (IoV), employing multiple public key generators for distributed registration. However, their scheme did not address scenarios where UAVs dynamically join or leave a cluster. Ali et al.~\cite{ref18} proposed a lightweight authentication scheme using resource-efficient cryptographic primitives. Yet, the scheme focused on mutual authentication between GBS and UAVs, making it inapplicable to cluster-based authentication. Wang et al.~\cite{ref19} presented a lightweight UAV authentication protocol, which similarly focused on mutual authentication rather than cluster authentication. Tanveer et al.~\cite{ref20} proposed a mutual authentication mechanism between UAVs and service users. Like~\cite{ref18,ref19}, it did not support either cluster or cross-cluster authentication.

\textbf{Current Limitations:} Although extensive research~\cite{ref13}--\cite{ref20} has explored authentication and key management for UAV networks, significant gaps remain in the context of dynamic UAV clusters. Most existing schemes~\cite{ref14},~\cite{ref17}--\cite{ref20} do not support scalable and efficient cluster authentication, particularly for simultaneous onboarding of multiple NUAVs. Moreover, privacy-preserving cross-cluster authentication remains largely unaddressed. Several schemes fail to ensure anonymity or unlinkability, thereby exposing EUAVs to flight trajectory inference attacks. In terms of session key management, many works either overlook backward secrecy or rely on computationly expensive operations, especially bilinear pairings~\cite{ref17}, which renders them impractical for resource-constrained UAV platforms. Furthermore, blockchain-based solutions~\cite{ref13,ref14,ref16} impose excessive computation and communication overheads, making them unsuitable for real-time, latency-sensitive applications in low-altitude economy environments.

These limitations hinder the reliability of UAV cluster services, thereby constraining their deployment in real-world low-altitude economic scenarios. To overcome these challenges, we propose LP2-CASKU that ensures reliable UAV cluster service delivery.

\begin{table*}[!t]
\scriptsize
\renewcommand{\arraystretch}{1.2}
\begin{threeparttable}
\caption{Comparison of Related Works}
\label{table1}
\centering
\begin{tabular}{p{2.4 cm} p{3.3 cm} p{2.3 cm} p{2 cm} p{2.9 cm} p{2.7cm}}
\hline
\textbf{Ref.} & \textbf{Main cryptographic primitives} & \textbf{Cluster authentication} & \textbf{Cross-cluster authentication} & \textbf{Cluster session key update} & \textbf{Privacy protection} \\
\hline
BDLA+~\cite{ref13} 2022 & ECC/Hash\textsuperscript{a} & $\checkmark$ & $\times$ & $\times$ & Anonymity \\
BCDA+~\cite{ref14} 2022 & ECC/SEA/Hash\textsuperscript{b} & $\times$ & $\checkmark$ & $\times$ & Anonymity \\
TAGKA~\cite{ref15} 2023 & CM/SS/Hash\textsuperscript{c} & $\checkmark$ & $\times$ & $\checkmark$ & Anonymity \\
SwarmAuth~\cite{ref16} 2024 & Hash/XOR/SEA & $\checkmark$ & $\checkmark$ & $\checkmark$ & Anonymity \\
BASUV~\cite{ref17} 2024 & BM/Hash & $\times$ & $\times$ & $\times$ & Anonymity \\
IOOSC-U2G~\cite{ref18} 2024 & ECC/Hash/XOR & $\times$ & $\times$ & $\times$ & Anonymity \\
LBMA+~\cite{ref19} 2024 & ECC/Hash/XOR & $\times$ & $\times$ & $\times$ & Anonymity \\
SAAF-IoD+~\cite{ref20} 2024 & CM/SEA/Hash & $\times$ & $\times$ & $\times$ & Anonymity \\
LP2-CASKU 2025 & Elgamal/Hash & $\checkmark$ & $\checkmark$ & $\checkmark$ & Anonymity/Unlinkability \\
\hline
\end{tabular}
\vspace{0.2cm}

\begin{tablenotes}
\footnotesize
\item[a-c] ECC denotes elliptic curve cryptography, SEA denotes symmetric encryption algorithm, CM denotes chaotic map, and SS denotes secret sharing.
\end{tablenotes}
\end{threeparttable}

\end{table*}

\section{Preliminaries and System Description}
The background knowledge and the system description are introduced in this section.
\subsection{Background Knowledge}
This section presents computationly hard problems, on which cryptograph primitives used in LP2-CASKU are based.

\textbf{Discrete Logarithm Problem (DLP)}~\cite{ref22}: Given a large prime $p$, a generator of the multiplicative group $\mathbb{Z}_{p}^{*}$ modulo $p$, and $b \in \mathbb{Z}_{p}^{*}$, it is computationly hard for any polynomial-time bounded algorithm to find $a \in \mathbb{Z}_{p}^{*}$ so that $b = g^{a}$.

\textbf{Diffie-Hellman Problem (DHP)}~\cite{ref23}: Given a large prime $p$, a generator of the multiplicative group $\mathbb{Z}_{p}^{*}$ modulo $p$, and $g^{a}, g^{b} \in \mathbb{Z}_{p}^{*}$, it is computationly hard for any polynomial-time bounded algorithm to compute $g^{ab} \in \mathbb{Z}_{p}^{*}$.

\subsection{System Description}
This section presents system entities, the threat model, design goals, and the security model. Table~\ref{table2} gives the symbols used in the rest of this paper.

\begin{table}[!t]
\renewcommand{\arraystretch}{1.2}
\caption{Definitions of Symbols}
\label{table2}
\centering
\begin{tabular}{p{1.5cm} p{6.4cm}}
\hline
\textbf{Symbol} & \textbf{Description} \\
\hline
$c_{\mathrm{NUAVs}}$ & Aggregated auxiliary information of $\{V_{k}\}_{k=1}^{N_{\mathrm{NUAV}}}$ \\
$\mathrm{CH}_{i,j}$ & The $j$-th CH managed by $\mathrm{GBS}_{i}$ \\
$\mathrm{CJT}_{i,j}$ & Cluster joining token of the cluster managed by $\mathrm{CH}_{i,j}$ \\
$\mathrm{CM}_{i,j,l}$ & The $l$-th CM in the cluster managed by $\mathrm{CH}_{i,j}$ \\
$\mathrm{CT}$ & Cross-cluster communication token \\
$g$ & Generator of $\mathbb{Z}_{p}^{*}$ \\
$\mathrm{GBS}_{i}$ & The $i$-th GBS \\
$\mathrm{H}$ & Hash function \\
$\mathrm{key}_{i,j}$ & Cluster session key of the cluster managed by $\mathrm{CH}_{i,j}$ \\
$\mathrm{key}_{i,j}^{\mathrm{new}}$ & Updated cluster session key of the cluster managed by $\mathrm{CH}_{i,j}$ \\

$N_{\mathrm{XX}}$ & Number of entity $\mathrm{XX}$ ($\mathrm{XX} \in \{\mathrm{GBS},  \mathrm{NUAV}, \mathrm{CH},$ $ \mathrm{CM}\}$) \\
$\mathrm{pk}_{\mathrm{CMs}}$ & Aggregated public key of $\{\mathrm{pk}_{\mathrm{CM}_{i,j,l}}\}_{l=1}^{N_{i,j,\mathrm{CM}}}$ \\
$(\mathrm{pk}_{\mathrm{XX}}, \mathrm{sk}_{\mathrm{XX}})$ & Public-private key pair of entity $\mathrm{XX}$ ($\mathrm{XX} \in \{\mathrm{GBS}, \mathrm{NUAV}, \mathrm{CH}, \mathrm{CM}, \mathrm{EUAV}\}$) \\
$\mathrm{pp}$ & Public parameters \\
$\mathrm{PID}_{\mathrm{XX}}$ & Pseudonymous identity of entity $\mathrm{XX}$ ($\mathrm{XX} \in \{\mathrm{NUAV}, \mathrm{CH}, \mathrm{CM}, \mathrm{EUAV}\}$) \\
$\mathrm{req}_{\mathrm{reg}}$ & Registration request message \\
$\mathrm{result}_{i,j,l}$ & Authentication result obtained by $\mathrm{CM}_{i,j,l}$ \\
$r_{\mathrm{XX}}$ & Random number for pseudonymous identity generation of $\mathrm{XX}$ ($\mathrm{XX} \in \{\mathrm{NUAV}, \mathrm{CM}\}$) \\
$\mathrm{RID}_{\mathrm{XX}}$ & Real identity of entity $\mathrm{XX}$ ($\mathrm{XX} \in \{\mathrm{NUAV}, \mathrm{CH}, $ $ \mathrm{CM}\}$) \\
$s_{i,j,l}$ & Random number used by $\mathrm{CM}_{i,j,l}$ \\
$\mathrm{sig}_{k}$ & Signature generated by $\mathrm{NUAV}_{k}$ for joining the cluster \\
$\mathrm{sig}_{\mathrm{CM}_{i,j,l}}$ & Signature generated by $\mathrm{CM}_{i,j,l}$ \\
$\mathrm{sig}_{\mathrm{CMs}}$ & Aggregated signature of $\{\mathrm{sig}_{\mathrm{CM}_{i,j,l}}\}_{l=1}^{N_{i,j,\mathrm{CM}}}$ \\
$\mathrm{sig}_{\mathrm{NUAVs}}$ & Aggregated signature of $\{\mathrm{sig}_{k}\}_{k=1}^{N_{\mathrm{NUAV}}}$ \\
$\mathrm{T}_{1}, \mathrm{T}_{2}, \mathrm{T}_{3}, \mathrm{T}_{4}$ & Timestamps \\
$u_{i,j}, \ u_{i,j,l}$ & Random numbers for cluster session key update \\
$v_{k}$ & Random number selected by $\mathrm{NUAV}_{k}$ \\
$V_{k}$ & Auxiliary information generated by $\mathrm{NUAV}_{k}$ for signature verification \\
$\mathbb{Z}_{p}^{*}$ & Multiplicative group modulo a large prime $p$ \\
$\lambda$ & Security parameter \\
$\oplus$ & Exclusive OR operation\\
\hline
\end{tabular}
\end{table}

\subsubsection{System Entity}
\begin{figure*}[!t]
\includegraphics[width=\textwidth]{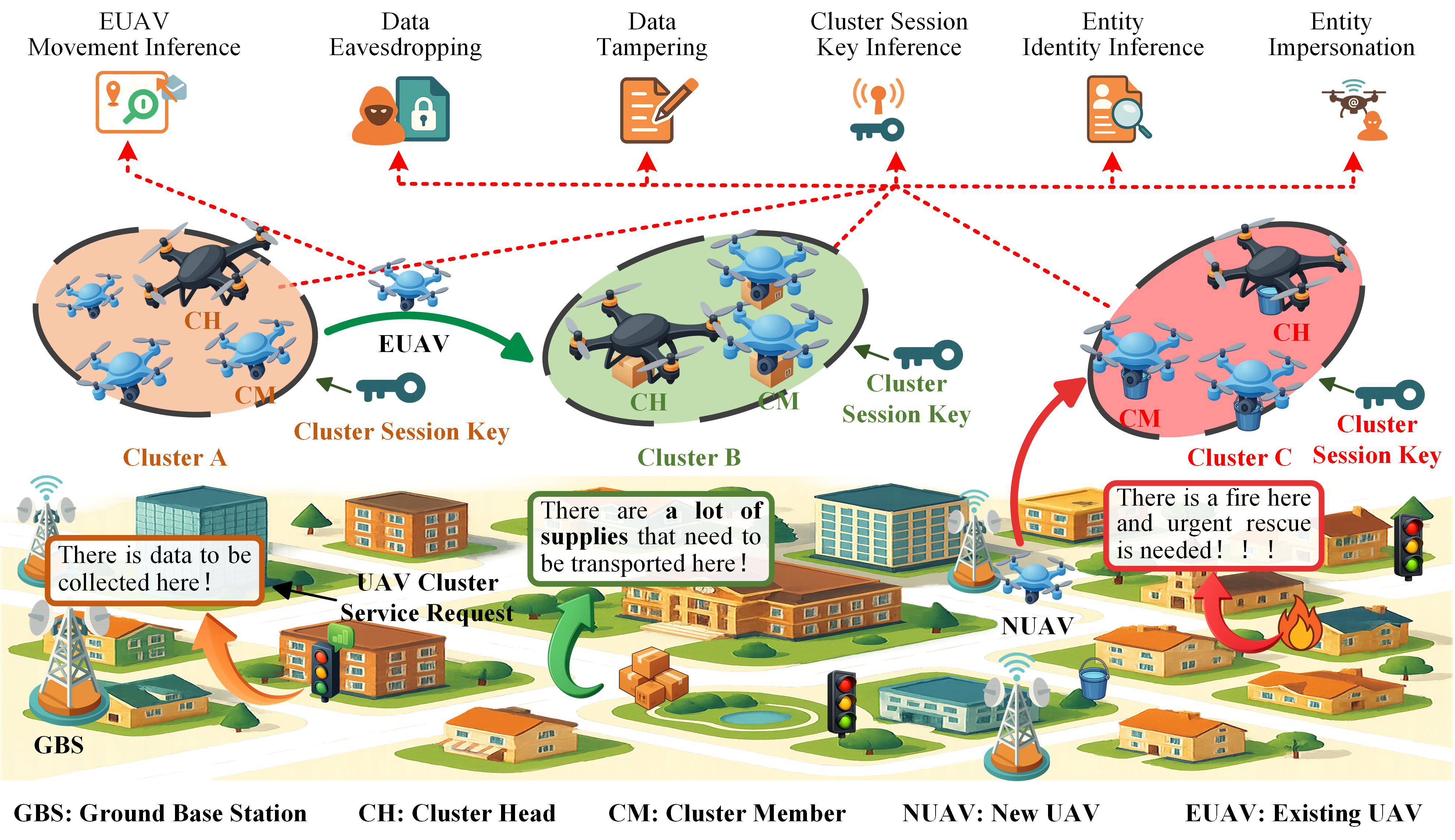}
\caption{Illustration of UAV cluster operations and potential security threats in low-altitude economy networks. Multiple UAV clusters dynamically provide services. Text boxes denote various UAV cluster service requests, including such as data collection, logistics transportation, and emergency rescue within the urban environment. Red dashed arrows indicate potential attack paths, each corresponding to specific attacks such as Data Eavesdropping, Data Tampering, Cluster Session Key Inference, EUAV Movement Inference, Entity Identity Inference, and Entity Impersonation mentioned in Section 3.2.2. The proposed LP2-CASKU addresses these attacks by enabling secure, privacy-preserving authentication and efficient UAV cluster management.}
\label{fig7}
\vspace{-10pt}
\end{figure*}

As illustrated in Fig.~\ref{fig7}, the proposed system involves five key entities: GBSs, CHs, CMs, NUAVs, and EUAVs. In the following, we provide a detailed description of these entities—not merely to define their roles, but to explain how each contributes to the reliability of UAV cluster services in the proposed scheme. Specifically, we describe the responsibilities of each entity in managing or participating in dynamic cluster operations, as well as their interaction with other entities to enable secure authentication, session key update, and coordinated task execution.

\textbf{GBS:} Each GBS oversees multiple UAV clusters within its designated swarm and is assumed to be fully trusted by all associated clusters. It is responsible for cluster formation, CH assignment, UAV registration, and global coordination. Upon receiving a request for reinforcement, the GBS registers NUAVs and dispatches them to the requesting cluster. Each GBS maintains a secure database containing identity credentials of all UAVs within the swarm.

\textbf{CH:} The CH is randomly selected by the GBS to serve as the coordinator of its respective cluster and is trusted by all CMs. Upon registration, the CH obtains a public-private key pair and a cluster session key from the GBS, enabling secure intra-cluster communication. Additionally, the CH receives a cross-cluster communication token, collaboratively generated by all GBSs, which authorizes secure interactions with CHs from other clusters, regardless of GBS affiliation. When additional UAVs are required, the CH may request NUAVs from its GBS or EUAVs from other CHs. All incoming NUAVs and EUAVs must be authenticated by the CH prior to integration into the cluster.

\textbf{CM:} CMs are UAVs managed under a common CH within the same cluster. They register with the GBS to obtain individual public-private key pairs and the cluster session key, which enable secure communication and coordinated task execution.

\textbf{NUAV:} NUAVs refer to UAVs newly introduced by the GBS to enhance the capabilities of a specific cluster. After registration, each NUAV obtains a public-private key pair and a cluster joining token. The NUAV initiates communication with the destination cluster's CH, and upon successful authentication by both the CH and its CMs, it receives the cluster session key and begins participating in coordinated operations.

\textbf{EUAV:} EUAVs are UAVs dynamically reassigned from one cluster to another upon the request of the destination cluster's CH. The source and destination clusters may operate under different GBSs. A EUAV may originate either from a CM that was part of the initial cluster formation or from a NUAV that joined the cluster during dynamic expansion. Prior to integration, the EUAV is authenticated by the destination CH using identity credentials retrieved from the GBS database, thereby completing the cross-cluster authentication process. Before initiating the transfer, the EUAV uploads its current state and mission-related data to the corresponding GBS via the source cluster's CH. To preserve both forward and backward secrecy, the cluster session keys of both the source and destination clusters are updated upon successful authentication and migration of the EUAV.

\subsubsection{Threat Model}

In this section, we examine potential security threats faced during UAV cluster service provision. Given that UAVs communicate over open wireless channels, the system is inherently susceptible to various attacks, including but not limited to data eavesdropping, data tampering, and entity impersonation. To systematically analyze these threats, we adopt the Dolev-Yao adversarial model~\cite{ref24}, which assumes that attackers have complete control over the communication channel, including the ability to intercept, modify, and fabricate messages. It is worth noting that denial-of-service attacks are beyond the scope of this work. The attacks considered in our model are illustrated in Fig.~\ref{fig7} and are described in detail as follows:

\begin{itemize}\setlength{\itemsep}{0pt}
    \item \textbf{Entity impersonation attack:} An adversary attempts to impersonate a legitimate system entity (e.g., CH, CM, EUAV, or NUAV) to gain unauthorized access or disrupt system operations during the dynamic changes of the UAV cluster~\cite{wang2024uavsurvey},~\cite{ceviz2024survey}.

    \item \textbf{Entity identity inference attack:} By analyzing communication content and patterns, particularly during the cluster authentication, an adversary seeks to deduce the true identity of participating entities.

    \item \textbf{EUAV movement inference attack:} Through correlating pseudonymous identifiers across multiple sessions, an adversary may infer the mobility trajectory of a EUAV, thereby violating its location privacy.

    \item \textbf{Data eavesdropping attack:} Sensitive information transmitted during the authentication and joining processes may be intercepted by an adversary monitoring the communication channel.

    \item \textbf{Data tampering attack:} An adversary actively modifies the content of messages exchanged during NUAV or EUAV authentication and integration, potentially compromising system integrity.

    \item \textbf{Cluster session key inference attack:} By exploiting intercepted protocol messages, an adversary attempts to infer the current or historical cluster session keys, thereby undermining the confidentiality of intra-cluster communications.
\end{itemize}

\subsubsection{Design Goals}
This section presents the design goals of LP2-CASKU, encompassing both security and performance considerations. The security goals aim to ensure the reliability of dynamic UAV cluster services in adversarial environments. In contrast, the performance goals address the stringent real-time requirements inherent to low-altitude economy networks.

\textbf{Security Goals:} To achieve secure UAV cluster operations, the proposed scheme is designed to fulfill the following security goals, which are analyzed in Section~5:

\begin{itemize}\setlength{\itemsep}{0pt}
    \item \textbf{S1) Authenticity of NUAVs and EUAVs:} The authenticity of NUAVs must be verified by both CH and CM before the integration of the cluster. Similarly, the authenticity of EUAVs must be validated by the destination CH during cross-cluster joining.

    \item \textbf{S2) Anonymity:} The true identities of CHs, CMs, NUAVs, and EUAVs must remain concealed from adversaries throughout the communication process.

    \item \textbf{S3) Unlinkability:} Messages generated by a EUAV across different cross-cluster authentication processes must not be linkable by adversaries, thereby preventing mobility trajectory inference.

    \item \textbf{S4) Enhanced forward secrecy:} Attackers cannot infer the new cluster session key even if they know the previous cluster session key. Besides, when a EUAV leaves its previous cluster, it cannot infer the new cluster session key even if it knows the previous cluster session key.

    \item \textbf{S5) Enhanced backward secrecy:} Attackers cannot infer the previous cluster session key even if they know the present cluster session key. Besides, when a NUAV or EUAV join a new cluster, it cannot infer the previous cluster session key even if it knows the present cluster session key.

    \item \textbf{S6) Message unforgeability:} All communication messages exchanged during NUAV and EUAV joining procedures must be protected against unauthorized modification and forgery.

    \item \textbf{S7) Message confidentiality:} The confidentiality of sensitive messages exchanged during NUAV joining and EUAV cross-cluster processes must be ensured.
\end{itemize}

\textbf{Performance Goals:} To support real-time operation and scalability in dynamic UAV cluster environments, we design LP2-CASKU to meet the following performance goals, which are evaluated in Section~6.4:

\begin{itemize}\setlength{\itemsep}{0pt}
    \item \textbf{P1) Lightweight cross-cluster authentication:} The authentication overhead incurred when a EUAV joins a new cluster should be significantly lower than that of a NUAV, thereby enabling rapid UAV integration.

    \item \textbf{P2) Low communication overhead for multiple NUAVs authentication:} When $N$ NUAVs simultaneously request to join a cluster, the total communication overhead should remain substantially below $N$ times the cost of authenticating a single NUAV, ensuring scalability.
\end{itemize}

\subsubsection{Security Model}

Security goals \textbf{S6} and \textbf{S7} ensure the unforgeability and confidentiality of transmitted data, forming the foundation of the scheme's overall security. Their formal analysis enables a systematic evaluation of data transmission security throughout the scheme. Therefore, we define the security model and conduct the formal analysis of \textbf{S6} and \textbf{S7} in this section. To be specific, we define cryptographic games~\cite{ref25} to simulate the interaction between entities in LP2-CASKU. Based on defined cryptographic games, we utilize a formal analysis tool~\cite{ref26} to conduct a comprehensive analysis of the vulnerabilities that exist in the system. Different security goals involve different interactions between entities and thus need to define different cryptographic games. Therefore, we define the data unforgeability game (DUG) and the data confidentiality game (DCG) for \textbf{S6} and \textbf{S7}, respectively.

\textbf{Data Unforgeability Game:} In DUG, there are a series of queries, used to simulate entity interactions, sent by a probabilistic polynomial time (PPT) adversary $\mathcal{A}_{1}$ to a challenger $\mathcal{C}$. DUG is detailed as follows:

\begin{itemize}[label=\textbullet]
    \item \textbf{Init:} $\mathcal{A}_{1}$ selects a set of challenge pseudonymous identities and declares the set.

    \item \textbf{Setup:} Given the security parameter $\lambda$, $\mathcal{C}$ generates the public parameters, which can be denoted as follows:
    \begin{equation}
        \mathrm{pp} = \{ p, g, \mathrm{pk}_{\mathrm{GBS}}, \mathrm{H} \}.
    \end{equation}

   \item \textbf{Query Phase:} $\mathcal{A}_{1}$ is allowed to issue the following types of queries to $\mathcal{C}$:

    \begin{itemize}[label=\textbullet]
        \item \textbf{Registration query:} Upon receiving a registration identity $\mathrm{RID}_{\mathcal{A}_{1}}$ from $\mathcal{A}_{1}$, $\mathcal{C}$ proceeds as follows based on the role specified. If $\mathrm{RID}_{\mathcal{A}_{1}} = \mathrm{RID}_{\mathrm{CH}}$, $\mathcal{C}$ responds with the message $\{\mathrm{PID}_{\mathcal{A}_{1}}, \mathrm{key}, \mathrm{sk}_{\mathcal{A}_{1}}, \mathrm{pk}_{\mathcal{A}_{1}}, \mathrm{CT}, \mathrm{CJT}\}$ by randomly selecting $\mathrm{key}, r_{\mathcal{A}_{1}}, \mathrm{CT}, \mathrm{CJT} \leftarrow \mathbb{Z}_{p}^{*}$ and the remaining values are computed as follows:
            \begin{equation}
            \left\{
            \begin{aligned}
                \mathrm{sk}_{\mathcal{A}_{1}} &= r_{\mathcal{A}_{1}} + \mathrm{sk}_{\mathrm{GBS}} \cdot \mathrm{H}(\mathrm{CJT}), \\
                \mathrm{pk}_{\mathcal{A}_{1}} &= g^{r_{\mathcal{A}_{1}}}, \\
                \mathrm{PID}_{\mathcal{A}_{1}} &= \mathrm{H}(\mathrm{sk}_{\mathcal{A}_{1}}, r_{\mathcal{A}_{1}}).
            \end{aligned}
            \right.
            \end{equation}
    
       If $\mathrm{RID}_{\mathcal{A}_{1}} = \mathrm{RID}_{\mathrm{CM}}$, $\mathcal{C}$ responds with the message $\{\mathrm{PID}_{\mathcal{A}_{1}}, \mathrm{key}, \mathrm{sk}_{\mathcal{A}_{1}}, \mathrm{pk}_{\mathcal{A}_{1}}\}$ by randomly selecting $r_{\mathcal{A}_{1}}, \mathrm{sk}_{\mathcal{A}_{1}}, \mathrm{key} \leftarrow \mathbb{Z}_{p}^{*}$ and the remaining values are computed as follows:
            \begin{equation}
            \left\{
            \begin{aligned}
                \mathrm{pk}_{\mathcal{A}_{1}} &= g^{\mathrm{sk}_{\mathcal{A}_{1}}}, \\
                \mathrm{PID}_{\mathcal{A}_{1}} &= \mathrm{H}(\mathrm{sk}_{\mathcal{A}_{1}}, r_{\mathcal{A}_{1}}).
            \end{aligned}
            \right.
            \end{equation}
    
        If $\mathrm{RID}_{\mathcal{A}_{1}} = \mathrm{RID}_{\mathrm{NUAV}}$, $\mathcal{C}$ responds with the message $\{\mathrm{PID}_{\mathcal{A}_{1}}, \mathrm{H}(\mathrm{CJT}), \mathrm{sk}_{\mathcal{A}_{1}}, \mathrm{pk}_{\mathcal{A}_{1}}\}$ by randomly selecting $\mathrm{sk}_{\mathcal{A}_{1}}, r_{\mathcal{A}_{1}}, \mathrm{key}, \mathrm{CJT} \leftarrow \mathbb{Z}_{p}^{*}$ and the the remaining values are computed as follows:
        \begin{equation}
        \left\{
        \begin{aligned}
            \mathrm{pk}_{\mathcal{A}_{1}} &= g^{\mathrm{sk}_{\mathcal{A}_{1}}}, \\
            \mathrm{PID}_{\mathcal{A}_{1}} &= \mathrm{H}(\mathrm{sk}_{\mathcal{A}_{1}}, r_{\mathcal{A}_{1}}).
        \end{aligned}
        \right.
        \end{equation}

        \item \textbf{Joining cluster request query:} Upon receiving the query from $\mathcal{A}_{1}$, $\mathcal{C}$ responds with the message $\{\mathrm{PID}_{\mathcal{A}_{1}}, \mathrm{PID}_{\mathrm{CH}}, \mathrm{sig}_{\mathcal{A}_{1}}, V_{\mathcal{A}_{1}}, \mathrm{pk}_{\mathcal{A}_{1}}\}$. Among them, $V_{\mathcal{A}_{1}}$ and $\mathrm{sig}_{\mathcal{A}_{1}}$ are computed as follows:
        \begin{equation}
        \left\{
        \begin{aligned}
            V_{\mathcal{A}_{1}} &= g^{v_{\mathcal{A}_{1}}}, \\
            \mathrm{sig}_{\mathcal{A}_{1}} &= D_{\mathcal{A}_{1}}^{v_{\mathcal{A}_{1}} \cdot w_{\mathcal{A}_{1}}},
        \end{aligned}
        \right.
        \end{equation}
        
        where $v_{\mathcal{A}_{1}}$ is randomly selected from $\mathbb{Z}_{p}^{*}$ and $D_{\mathcal{A}_{1}}$ and $w_{\mathcal{A}_{1}}$ are intermediate values computed as follows:
        
        \begin{equation}
        \left\{
        \begin{aligned}
            D_{\mathcal{A}_{1}} &= \mathrm{pk}_{\mathrm{GBS}}^{\mathrm{H}(\mathrm{CJT})} \cdot \mathrm{pk}_{\mathrm{CH}}, \\
            w_{\mathcal{A}_{1}} &= \mathrm{H}(\mathrm{PID}_{\mathcal{A}_{1}}, \mathrm{PID}_{\mathrm{CH}}, \mathrm{pk}_{\mathcal{A}_{1}}).
        \end{aligned}
        \right.
        \end{equation}

        \item \textbf{CM verification query:} $\mathcal{A}_{1}$ sends this query to $\mathcal{C}$. $\mathcal{C}$ responds with the message $\{\mathrm{sig}_{\mathcal{A}_{1}}, c_{\mathcal{A}_{1}}, \mathrm{T}_{1}\}$. Among them, $\mathrm{T}_{1}$ is a timestamp selected by $\mathcal{C}$ and  $\mathrm{sig}_{\mathcal{A}_{1}}$ and $c_{\mathcal{A}_{1}}$ are computed as follows:
        \begin{equation}
        \left\{
        \begin{aligned}
            \mathrm{sig}_{\mathcal{A}_{1}} &= g^{(\mathrm{H}(\mathrm{result}) - \mathrm{sk}_{\mathcal{A}_{1}} \cdot M)s^{-1}}, \\
            c_{\mathcal{A}_{1}} &= \mathrm{result} \oplus \mathrm{key},
        \end{aligned}
        \right.
        \end{equation}
        where $\mathrm{result}$ is an intermediate values computed as follows:
        \begin{equation}
        \left\{
        \begin{aligned}
            \mathrm{result} &= \mathrm{H}(\mathrm{T}_{1}, \mathrm{key}).
        \end{aligned}
        \right.
        \end{equation}

        \item \textbf{CH verification query:} $\mathcal{A}_{1}$ sends this query to $\mathcal{C}$. $\mathcal{C}$ responds with the message $\{Q_{\mathcal{A}_{1}}, \mathrm{T}_{2}\}$, where $\mathrm{T}_{2}$ is a timestamp selected by $\mathcal{C}$ and $Q_{\mathcal{A}_{1}}$ is computed as follows:
        \begin{equation}
            Q_{\mathcal{A}_{1}} = \mathrm{H}(\mathrm{result}', \mathrm{T}_{2}).
        \end{equation}

        \item \textbf{Cross-cluster verification query:} $\mathcal{A}_{1}$ sends this query to $\mathcal{C}$. $\mathcal{C}$ responds with the message $\{C_{\mathrm{CH}}, \mathrm{PID}_{\mathcal{A}_{1}}, \mathrm{T}_{3}\}$, where $\mathrm{T}_{3}$ is a timestamp selected by $\mathcal{C}$ and $C_{\mathrm{CH}}$ is computed as follows:
        \begin{equation}
            C_{\mathrm{CH}} = \mathrm{H}(\mathrm{PID}_{\mathcal{A}_{1}}, \mathrm{T}_{3}, \mathrm{CT}) \oplus \mathrm{CT}.
        \end{equation}
    \end{itemize}

    \item \textbf{Challenge:} After the \textbf{Query Phase}, $\mathcal{A}_{1}$ is provided with the message $\{ \mathrm{PID}_{\mathcal{A}_{1}}, \mathrm{PID}_{\mathrm{CH}}, \mathrm{pk}_{\mathcal{A}_{1}}, \mathrm{pk}_{\mathrm{CH}}, T_{1}, T_{2}, T_{3} \}$ by $\mathcal{C}$, none of which were requested during the \textbf{Query Phase}.

    \item \textbf{Guess:} If $\mathcal{A}_{1}$ can generate any valid value among $(\mathrm{sig}_{\mathcal{A}_{1}}, V_{\mathcal{A}_{1}})$, $(\mathrm{sig}_{\mathcal{A}_{1}}, c_{\mathcal{A}_{1}})$, $Q_{\mathcal{A}_{1}}$, or $C_{\mathcal{A}_{1}}$, then $\mathcal{A}_{1}$ wins DUG.
\end{itemize}

\textbf{Data Confidentiality Game:} DCG also has a series of queries, used to simulate entity interactions, sent by a PPT adversary $\mathcal{A}_{2}$ to a challenger $\mathcal{C}$. DUG is detailed as follows:
\begin{itemize}[label=\textbullet]
  \item \textbf{Init:} $\mathcal{A}_{2}$ selects a set of challenge pseudonymous identities and declares the set.

  \item \textbf{Setup:} Given the security parameter $\lambda$, $\mathcal{C}$ generates the public parameters, which can be denoted as follows:
  \begin{equation}
    \mathrm{pp} = \{ p, g, \mathrm{pk}_{\mathrm{GBS}}, \mathrm{H} \}.
  \end{equation}

  \item \textbf{Query Phase:} In addition to the registration, CM verification, and cross-cluster verification queries defined in DUG, DCG introduces the following additional query:

  \begin{itemize}[label=\textbullet]
    \item \textbf{Cluster session key update query:} Upon receiving a message $m$ from $\mathcal{A}_{2}$, $\mathcal{C}$ randomly selects a new session key $\mathrm{key}^{\mathrm{new}} \leftarrow \mathbb{Z}_{p}^{*}$ and constructs a polynomial $f(x)$ of degree $N_{\mathrm{CM}} - 1$, where $f(x)$ is denoted as follows:
    \begin{equation}
      f(x) = \mathrm{key}^{\mathrm{new}} + b_1 x + \cdots + b_{N_{\mathrm{CM}} - 1} x^{N_{\mathrm{CM}} - 1}.
    \end{equation}

    Then $\mathcal{C}$ responds with the message $\{ \mathrm {key}^{\mathrm{new}} \oplus m \}$, where $\mathrm{key} ^{\mathrm{new}}$ is computed as follows:
    \begin{equation}
    \mathrm{key}^{\mathrm{new}} = \sum\nolimits_{l = 1}^{N_{\mathrm{CM}}} \left( \mathrm{key}_l^{\mathrm{new}} \prod\nolimits_{n = 1, n \ne l}^{N_{\mathrm{CM}}} \left( - \frac{x_n}{x_l - x_n} \right) \right).
    \label{13}
    \end{equation}
    The value $\mathrm{key}_l^{\mathrm{new}}$ in Eq.~(\ref{13}) in computed as follows:
    \begin{equation}
      \mathrm{key}_l^{\mathrm{new}} = U_l \oplus g^{f(x_l) f(x_n)},
    \end{equation}
    where $x_l$, $f(x_l)$, and $U_l$ are computed as follows:
    \begin{equation}
    \left\{
    \begin{aligned}
        x_l &= \mathrm{H}(\mathrm{PID}_{\mathrm{CM}_l}), \\
        f(x_l) &= \mathrm{key}^{\mathrm{new}} + b_1 x_l + \cdots + b_{N_{\mathrm{CM}} - 1} x_l^{N_{\mathrm{CM}} - 1}, \\
        U_l &= g^{f(x_n) f(x_l)} \oplus f(x_l).
    \end{aligned}
    \right.
    \end{equation}
  \end{itemize}

  \item \textbf{Challenge:} After the \textbf{Query Phase}, $\mathcal{A}_{2}$ submits two challenge messages $\{ m_0, m_1 \}$ that have not appeared in previous queries. $\mathcal{C}$ randomly selects $b \leftarrow \{ 0, 1 \}$ and responds with the message $\{ c_b \}$, where $c_b$ is computed as follows: 
  \begin{equation}
    c_b = \mathrm{Enc}(\mathrm{key}^{\mathrm{new}}, m_b).
  \end{equation}

  \item \textbf{Guess:} Upon receiving $c_b$, $\mathcal{A}_{2}$ outputs a guess $b' \in \{ 0, 1 \}$. If $b' = b$, then $\mathcal{A}_{2}$ successfully wins DCG.
\end{itemize}

\section{Construction of LP2-CASKU}

This section details the construction of LP2-CASKU. Before given the detailed description about LP2-CASKU, we firstly introduce  the main idea about why and how we design it.

\subsection{Main Idea}

To ensure reliable UAV cluster services in highly dynamic and resource-constrained environments, it is imperative to address several critical challenges, including efficient scalable cluster authentication, privacy-preserving cross-cluster authentication, and secure cluster session key management. Existing schemes often fail to simultaneously satisfy these requirements, particularly under the stringent latency and efficiency constraints imposed by low-altitude economy networks. To bridge this gap, we propose LP2-CASKU, which comprises three core mechanisms: a message aggregation mechanism (\textbf{MAm}), a lightweight cross-cluster authentication mechanism (\textbf{LC2Am}), and a cluster session key update mechanism (\textbf{CSKUm}).

LP2-CASKU guarantees the authenticity of dynamically joining UAVs, including both NUAVs and EUAVs, thereby preventing adversaries from forging UAV identities during cluster evolution. It also ensures the confidentiality and integrity of intra-cluster communications through secure session key updates, even in the presence of UAV joins and leaves. In addition, LP2-CASKU mitigates privacy risks associated with EUAV mobility across clusters, by preserving anonymity and unlinkability. Furthermore, the proposed design significantly reduces the authentication and key update overhead introduced by frequent cluster changes, achieving lightweight performance without sacrificing security. These combined capabilities enable LP2-CASKU to effectively address the above three critical challenges. The detailed threat model and design goals are presented in Sections 3.2.2 and 3.

\textbf{Motivation:} The motivations behind the design of these mechanisms are outlined as follows:

\begin{itemize}
    \item \textbf{MAm:} In LP2-CASKU, when a NUAV joins a cluster, it must be authenticated by both the CH and all associated CMs. Moreover, since a NUAV may subsequently serve as a EUAV in other clusters, its authentication result must also be shared with other CHs. In scenarios involving the deployment of multiple NUAVs, performing individual authentication and communication for each NUAV would incur excessive overhead and latency. To address this, \textbf{MAm} is introduced to aggregate authentication requests and responses. Specifically, the CH aggregates multiple NUAV join requests and broadcasts them collectively to the CMs, rather than sending them individually. Similarly, authentication responses from CMs are aggregated and forwarded as a single result to other CHs. This significantly reduces latency, enabling efficient and scalable cluster authentication suitable for low-latency UAV networks.

    \item \textbf{LC2Am:} EUAVs are typically legitimate CMs that have already undergone initial authentication. To avoid redundant authentication and protect EUAV privacy, \textbf{LC2Am} leverages the principles of Single Sign-On (SSO)~\cite{ref27}. This mechanism ensures that EUAVs can be authenticated across clusters while maintaining both anonymity and unlinkability, thus preserving privacy without sacrificing efficiency.

    \item \textbf{CSKUm:} When a UAV dynamically joins or leaves a cluster—whether as a NUAV or EUAV—it is essential to update the cluster session key to preserve its secrecy. \textbf{CSKUm} is designed to ensure both forward secrecy (i.e., newly NUAV cannot access previous keys) and backward secrecy (i.e., departed UAVs cannot access future keys), thereby strengthening overall key confidentiality and system robustness.
\end{itemize}

By integrating these mechanisms, LP2-CASKU effectively addresses the challenges outlined in Section~1 and ensures the secure, efficient, and privacy-preserving operation of dynamic UAV clusters in low-altitude economy scenarios.

\subsection{Overall Process}
LP2-CASKU consists of five phases: Setup, Registration, Join, Cross-cluster, and Cluster Session Key Update.

\begin{itemize}
    \item \textbf{Setup Phase:} All $\mathrm{GBS}$s jointly generate the public parameters for system initialization.
    
    \item \textbf{Registration Phase:} Each CH and CM registers its identity via its associated GBS, which stores the verified identities for future use.
    
    \item \textbf{Join Phase:} When a cluster needs additional UAVs, the CH sends a request to the GBS, which assigns NUAVs. Mutual authentication between the NUAVs and the cluster is performed using \textbf{MAm}. Once authenticated, the NUAVs join the cluster.
    
    \item \textbf{Cross-cluster Phase:} A CH may recruit EUAVs from other clusters. Before integration, each EUAV is authenticated using the proposed \textbf{LC2Am}.
    
    \item \textbf{Cluster Session Key Update Phase:} The proposed \textbf{CSKUm} is invoked to update the cluster session key when UAVs join or leave, ensuring forward and backward secrecy.
\end{itemize}

\subsubsection{Setup Phase}
In this phase, all $\mathrm{GBS}$s jointly determine the public parameters. Assuming there are $N_{\mathrm{GBS}}$ GBSs, the detailed steps are as follows:

\begin{enumerate}[label=\textbf{\textit{Step \arabic*: }}, align=left, leftmargin=0pt, labelsep=0em]
    \item Given the security parameter $\lambda$ and a large prime $p$ with generator $g \in \mathbb{Z}_{p}^{*}$, each $\mathrm{GBS}_{i}$ randomly selects $\mathrm{sk}_{\mathrm{GBS}_{i}} \leftarrow \mathbb{Z}_{p}^{*}$ and computes its public key $\mathrm{pk}_{\mathrm{GBS}_{i}}$ as follows: 
    \begin{equation}
        \mathrm{pk}_{\mathrm{GBS}_{i}} = g^{\mathrm{sk}_{\mathrm{GBS}_{i}}}.
    \end{equation}
    Here, the security parameter $\lambda$ determines the bit length of the prime $p$, thereby controlling the cryptographic strength of the system~\cite{goldwasser1999}. A larger $\lambda$ provides a higher level of security by increasing the computational difficulty of solving discrete logarithm problems over $\mathbb{Z}_{p}^{*}$, at the cost of greater computational and communication overhead. We assume that GBSs are fully trusted and serve as a lightweight public key infrastructure, responsible for securely generating and distributing public keys to network entities. This assumption, common in cryptographic systems, allows each public key to be verifiably associated with the identity of its corresponding entity~\cite{albarqi2015pki}. Then all $\mathrm{GBS}$s jointly select a hash function $\mathrm{H}: \mathbb{Z}_{p}^{*} \to \mathbb{Z}_{p}^{*}$ and the cross-cluster communication token $\mathrm{CT} \leftarrow \mathbb{Z}_{p}^{*}$.

    \item Finally, all $\mathrm{GBS}$s publish the public parameters, which can be denoted as follows:
    \begin{equation}
        \mathrm{pp} = \{ p, g, \{ \mathrm{pk}_{\mathrm{GBS}_{i}} \}_{i=1}^{N_{\mathrm{GBS}}}, \mathrm{H} \}.
    \end{equation}
\end{enumerate}

\subsubsection{Registration Phase}

In this phase, each CH and CM register their identity information via the GBS they belong to. After that, the GBS stores their identity information in its database.

\textbf{CH Registration:} Assume there are $N_{i,\mathrm{CH}}$ clusters managed by $\mathrm{GBS}_{i}$. Taking $\mathrm{CH}_{i,j}$ ($j \in [1, N_{i,\mathrm{CH}}]$) as an example, the detailed CH registration process is as follows:

\begin{enumerate}[label=\textbf{\textit{Step \arabic*: }}, align=left, leftmargin=0pt, labelsep=0em]

    \item $\mathrm{CH}_{i,j}$ initiates the registration procedure by sending the message $\{ \mathrm{RID}_{\mathrm{CH}_{i,j}}, \mathrm{req}_{\mathrm{reg}} \}$ to $\mathrm{GBS}_{i}$ via a secure channel.

    \item Upon receiving the message, $\mathrm{GBS}_{i}$ returns the message $\{ \mathrm{CT}, \mathrm{key}_{i,j},  \mathrm{CJT}_{i,j}, \mathrm{sk}_{\mathrm{CH}_{i,j}}, \mathrm{pk}_{\mathrm{CH}_{i,j}}, \mathrm{PID}_{\mathrm{CH}_{i,j}} \}$ to $\mathrm{CH}_{i,j}$ over the secure channel. Among the message, $\mathrm{CT}$ is obtained in the Setup Phase, $\mathrm{key}_{i,j}, \mathrm{CJT}_{i,j}$ are selected from $\mathbb{Z}_{p}^{*}$, and other values are computed as follows:
    \begin{equation}
    \left\{
    \begin{aligned}
        \mathrm{sk}_{\mathrm{CH}_{i,j}} &= r_{\mathrm{CH}_{i,j}} + \mathrm{sk}_{\mathrm{GBS}_{i}} \cdot \mathrm{H}(\mathrm{CJT}_{i,j}), \\
        \mathrm{pk}_{\mathrm{CH}_{i,j}} &= g^{r_{\mathrm{CH}_{i,j}}}, \\
        \mathrm{PID}_{\mathrm{CH}_{i,j}} &= \mathrm{H}(\mathrm{sk}_{\mathrm{CH}_{i,j}}, r_{\mathrm{CH}_{i,j}}),
    \end{aligned}
    \right.
    \end{equation}
    where $r_{\mathrm{CH}_{i,j}}$ is also selected from $\mathbb{Z}_{p}^{*}$.

    \item Finally, $\mathrm{GBS}_{i}$ records $\mathrm{key}_{i,j}$ ,  $\mathrm{CJT}_{i,j}$ , and  $\mathrm{PID}_{\mathrm{CH}_{i,j}}$ in its local database and broadcasts the message $\{\mathrm{PID}_{\mathrm{CH}_{i,j}}\}$ to all other $\mathrm{GBS}$s, which also store $\mathrm{PID}_{\mathrm{CH}_{i,j}}$ locally for future authentication purposes.

\end{enumerate}

\textbf{CM Registration:} Assume there are $N_{i,j,\mathrm{CM}}$ CMs in the $j$-th cluster of $\mathrm{GBS}_{i}$. Taking $\mathrm{CM}_{i,j,l}$ ($l \in [1, N_{i,j,\mathrm{CM}}]$) as an example, the detailed $\mathrm{CM}$ registration process is as follows:

\begin{enumerate}[label=\textbf{\textit{Step \arabic*: }}, align=left, leftmargin=0pt, labelsep=0em]

    \item $\mathrm{CM}_{i,j,l}$ sends the message $\{ \mathrm{RID}_{\mathrm{CM}_{i,j,l}}, \mathrm{req}_{\mathrm{reg}} \}$ to $\mathrm{GBS}_{i}$ via a secure channel.
    
    \item Upon receiving the message, $\mathrm{GBS}_{i}$ retrieves $\mathrm{key}_{i,j}$ from its database and returns the message $\{ \mathrm{key}_{i,j}, \mathrm{sk}_{\mathrm{CM}_{i,j,l}}, \mathrm{pk}_{\mathrm{CM}_{i,j,l}}, \mathrm{PID}_{\mathrm{CM}_{i,j,l}} \}$ to $\mathrm{CM}_{i,j,l}$ via a secure channel. Among the message, $\mathrm{sk}_{\mathrm{CM}_{i,j,l}}$ is selected from $\mathbb{Z}_{p}^{*}$ and other values are computed as follows:
    \begin{equation}
    \left\{
    \begin{aligned}
        \mathrm{pk}_{\mathrm{CM}_{i,j,l}} &= g^{\mathrm{sk}_{\mathrm{CM}_{i,j,l}}}, \\
        \mathrm{PID}_{\mathrm{CM}_{i,j,l}} &= \mathrm{H}(\mathrm{sk}_{\mathrm{CM}_{i,j,l}}, r_{\mathrm{CM}_{i,j,l}}),
    \end{aligned}
    \right.
    \end{equation}
    where $r_{\mathrm{CM}_{i,j,l}}$ is also selected from $\mathbb{Z}_{p}^{*}$. It is worth noting that the CH and all associated CMs within the same UAV cluster share a common symmetric key $\mathrm{key}_{i,j}$.

    \item Finally, $\mathrm{GBS}_{i}$ stores $\mathrm{PID}_{\mathrm{CM}_{i,j,l}}$ in its database and broadcasts it to other $\mathrm{GBS}$s, which also store $\mathrm{PID}_{\mathrm{CM}_{i,j,l}}$ in their respective databases.

\end{enumerate}

Upon completing the registration process, each CH and its associated CMs form a UAV cluster to execute collaborative tasks. Each cluster consists of one CH and multiple CMs. Secure communication between the CH and its CMs is enabled via the shared symmetric key $\mathrm{key}_{i,j}$ established during registration.

\subsubsection{Join Phase with \textbf{MAm}}

In this phase, a NUAV joins an existing UAV cluster. Specifically, when a cluster requires additional UAVs to complete assigned tasks, its corresponding CH sends a request to the associated GBS. Upon receiving the request, the GBS initializes a set of NUAVs to serve as supplementary members. Subsequently,  the proposed \textbf{MAm} is used to facilitate efficient mutual authentication between the NUAVs and the cluster. Once authenticated, NUAVs are authorized to join the cluster and participate in task execution.

\textbf{Cluster Authentication:} Let $\mathrm{GBS}_{i}$ ($i \in [1, N_{\mathrm{GBS}}]$) denote the GBS, $\mathrm{CH}_{i,j}$ ($i \in [1, N_{\mathrm{GBS}}], j \in [1, N_{i,\mathrm{CH}}]$) denotes the CH of the cluster that NUAVs join, $\{ \mathrm{CM}_{i,j,l} \}_{l=1}^{N_{i,j,\mathrm{CM}}}$ denote the group of CMs managed by $\mathrm{CH}_{i,j}$, and $\{ \mathrm{NUAV}_{k} \}_{k=1}^{N_{\mathrm{NUAV}}}$ denote the group of NUAVs. The details of each step are as follows:

\begin{enumerate}[label=\textbf{\textit{Step \arabic*: }}, align=left, leftmargin=0pt, labelsep=0em]

    \item $\mathrm{GBS}_{i}$ sends the message $\{ \mathrm{H}(\mathrm{CJT}_{i,j}), \mathrm{pk}_{\mathrm{CH}_{i,j}},$ $\mathrm{PID}_{\mathrm{CH}_{i,j}}, \mathrm{sk}_{\mathrm{NUAV}_{k}}, \mathrm{pk}_{\mathrm{NUAV}_{k}},  \mathrm{PID}_{\mathrm{NUAV}_{k}} \}$ to the corresponding $\mathrm{NUAV}_{k}$ via a secure channel. Among the message, $\mathrm{CJT}_{i,j}$ is retrieved from the internal database of $\mathrm{GBS}_{i}$, $\mathrm{sk}_{\mathrm{NUAV}_{k}}$ is selected from $\mathbb{Z}_{p}^{*}$, $\mathrm{pk}_{\mathrm{CH}_{i,j}}$ and $\mathrm{PID}_{\mathrm{CH}_{i,j}}$ are the public key and PID of $\mathrm{CH}_{i,j}$, respectively, and the remaining value are obtained as shown in the following equation:
    \begin{equation}
    \left\{
    \begin{aligned}
        \mathrm{pk}_{\mathrm{NUAV}_{k}} &= g^{\mathrm{sk}_{\mathrm{NUAV}_{k}}}, \\
        \mathrm{PID}_{\mathrm{NUAV}_{k}} &= \mathrm{H}(\mathrm{sk}_{\mathrm{NUAV}_{k}}, r_{\mathrm{NUAV}_{k}}),
    \end{aligned}
    \right.
    \end{equation}
    where $r_{\mathrm{NUAV}_{k}}$ is also selected from $\mathbb{Z}_{p}^{*}$.

    \item Each $\mathrm{NUAV}_{k}$ sends the message $\{ \mathrm{PID}_{\mathrm{NUAV}_{k}},$ $ \mathrm{pk}_{\mathrm{NUAV}_{k}}, $ $\mathrm{PID}_{\mathrm{CH}_{i,j}}, V_{k}, \mathrm{sig}_{k} \}$ to $\mathrm{CH}_{i,j}$. Among the message, $\mathrm{PID}_{\mathrm{NUAV}_{k}}$ and $\mathrm{pk}_{\mathrm{NUAV}_{k}}$ are the public key and PID of $\mathrm{NUAV}_k$, respectively, $\mathrm{PID}_{\mathrm{CH}_{i,j}}$ is the PID of $\mathrm{CH}_{i,j}$, and other values are computed as follows:
    \begin{equation}
    \left\{
    \begin{aligned}
        V_{k} &= g^{v_{k}}, \\
        \mathrm{sig}_{k} &= D_{k}^{v_{k} w_{k}},
    \end{aligned}
    \right.
    \end{equation}
    where $v_{k}$ is selected from $\mathbb{Z}_{p}^{*}$ and $D_{k}$ and $w_{k}$ are computed as follows:
    \begin{equation}
    \left\{
    \begin{aligned}
        D_{k} &= \mathrm{pk}_{\mathrm{GBS}_{i}}^{\mathrm{H}(\mathrm{CJT}_{i,j})} \cdot \mathrm{pk}_{\mathrm{CH}_{i,j}}, \\
        w_{k} &= \mathrm{H}(\mathrm{PID}_{\mathrm{NUAV}_{k}}, \mathrm{PID}_{\mathrm{CH}_{i,j}}, \mathrm{pk}_{\mathrm{NUAV}_{k}}).
    \end{aligned}
    \right.
    \end{equation}

    \item After receiving messages from $\{ \mathrm{NUAV}_{k} \}_{k=1}^{N_{\mathrm{NUAV}}}$, $\mathrm{CH}_{i,j}$ sends the message $\{ \mathrm{PID}_{\mathrm{CH}_{i,j}}, \mathrm{T}_{1}, \mathrm{sig}_{\mathrm{NUAVs}}, c_{\mathrm{NUAVs}},$ $ S_{i,j,l}, M, K_{\mathrm{CM}_{i,j,l}} \}$ to each $\mathrm{CM}_{i,j,l}$. Among the message, $\mathrm{PID}_{\mathrm{CH}_{i,j}}$ is the PID of $\mathrm{CH}_{i,j}$, $\mathrm{T}_{1}$ is a timestamp, and other values are computed as follows:
    \begin{equation}
    \left\{
    \begin{aligned}
        \mathrm{sig}_{\mathrm{NUAVs}} &= \mathrm{H} \left( \left( \prod_{k=1}^{N_{\mathrm{NUAV}}} \mathrm{sig}_{k} \right)^{\mathrm{sk}_{\mathrm{CH}_{i,j}}^{-1}} \right) \oplus \mathrm{key}_{i,j}, \\
        c_{\mathrm{NUAVs}} &= \prod_{k=1}^{N_{\mathrm{NUAV}}} V_{k}^{\mathrm{H}(\mathrm{PID}_{\mathrm{NUAV}_{k}}, \mathrm{PID}_{\mathrm{CH}_{i,j}}, \mathrm{pk}_{\mathrm{NUAV}_{k}})}, \\
        S_{i,j,l} &= s_{i,j,l} \oplus \mathrm{H}(\mathrm{key}_{i,j}, \mathrm{T}_{1}), \\
        M &= \sum_{l=1}^{N_{i,j,\mathrm{CM}}} s_{i,j,l},, \\
        K_{\mathrm{CM}_{i,j,l}} &= \mathrm{H}(s_{i,j,l}, \mathrm{PID}_{\mathrm{CM}_{i,j,l}}, M),
    \end{aligned}
    \right.
    \end{equation}
    where $s_{i,j,l}$ is selected from $\mathbb{Z}_{p}^{*}$.

    \item Upon receiving the message from $\mathrm{CH}_{i,j}$, each $\mathrm{CM}_{i,j,l}$ first verifies the freshness of $\mathrm{T}_{1}$. If $\mathrm{T}_{1}$ is determined to be fresh, each $\mathrm{CM}_{i,j,l}$ proceeds to verify the following equation:
    \begin{equation}
    \mathrm{sig}_{\mathrm{NUAVs}} \oplus \mathrm{key}_{i,j} \stackrel{?}{=} \mathrm{H}(c_{\mathrm{NUAVs}}).
    \label{eq:48}
    \end{equation}
    If Eq.~(\ref{eq:48}) holds, each $\mathrm{CM}_{i,j,l}$ continues to validate the following equality:
    \begin{equation}
    K_{\mathrm{CM}_{i,j,l}} \stackrel{?}{=} \mathrm{H}(s_{i,j,l}', \mathrm{PID}_{\mathrm{CM}_{i,j,l}}, M),
    \label{eq:49}
    \end{equation}
    where $s_{i,j,l}'$ is a recovered share, computed as follows:
    \begin{equation}
    s_{i,j,l}' = S_{i,j,l} \oplus \mathrm{H}(\mathrm{key}_{i,j}, \mathrm{T}_{1}).
    \label{eq:50}
    \end{equation}
    If Eq.~(\ref{eq:49}) is satisfied, the shared result is computed as follows:
    \begin{equation}
    \mathrm{result} = \mathrm{H}(\mathrm{PID}_{\mathrm{CH}_{i,j}}, \mathrm{T}_{1}, \mathrm{key}_{i,j});
    \label{eq:51}
    \end{equation}
    otherwise, a fallback result is computed as follows:
    \begin{equation}
    \mathrm{result} = \mathrm{H}(\mathrm{T}_{1}, \mathrm{key}_{i,j}).
    \label{eq:52}
    \end{equation} 
    Finally, each $\mathrm{CM}_{i,j,l}$ sends the message $\{\mathrm{T}_{1}, $ $ \mathrm{sig}_{\mathrm{CM}_{i,j,l}}, $ $ c_{\mathrm{CM}_{i,j,l}}  \}$ back to $\mathrm{CH}_{i,j}$. Among the message, $\mathrm{T}_{1}$ is the timestamp received and the values $\mathrm{sig}_{\mathrm{CM}_{i,j,l}}$ and $c_{\mathrm{CM}_{i,j,l}}$ are computed as follows:
    \begin{equation}
    \left\{
    \begin{aligned}
    \mathrm{sig}_{\mathrm{CM}_{i,j,l}} &= g^{\left( N_{i,j,\mathrm{CM}} \cdot \mathrm{H}(\mathrm{result}) - \mathrm{sk}_{\mathrm{CM}_{i,j,l}} \cdot M \right) s_{i,j,l}^{-1}}, \\
    c_{\mathrm{CM}_{i,j,l}} &= \mathrm{result} \oplus \mathrm{key}_{i,j},
    \end{aligned}
    \right.
    \label{eq:53_54}
    \end{equation}
    where the result is computed from Eq.(\ref{eq:51}) if Eq.(\ref{eq:49}) is satisfied; otherwise, it is computed from Eq.~(\ref{eq:52}). In both cases, $M$ is obtained from $\mathrm{CH}_{i,j}$ and $\mathrm{sk}_{\mathrm{CM}_{i,j,l}}$ is the private key of $\mathrm{CM}_{i,j,l}$.

    \item Upon receiving messages from $\{ \mathrm{CM}_{i,j,l} \}_{l=1}^{N_{i,j,\mathrm{CM}}}$, if the timestamp $\mathrm{T}_{1}$ is verified as fresh, $\mathrm{CH}_{i,j}$ proceeds to compute the following equation:
    \begin{equation}
    \left\{
    \begin{aligned}
    \mathrm{result}_{i,j,l} &= c_{\mathrm{CM}_{i,j,l}} \oplus \mathrm{key}_{i,j}, \quad \forall l \in [1, N_{i,j,\mathrm{CM}}], \\
    \mathrm{result}_{i,j,1} &= \mathrm{H}(\mathrm{PID}_{\mathrm{CH}_{i,j}}, \mathrm{T}_{1}, \mathrm{key}_{i,j}).
    \end{aligned}
    \right.
    \end{equation}
    Then $\mathrm{CH}_{i,j}$ verifies the following equation:        \begin{equation}
    \mathrm{result}_{i,j,1} \stackrel{?}{=} \cdots \stackrel{?}{=} \mathrm{result}_{i,j,N_{i,j,\mathrm{CM}}}.
    \label{eq:55_57}
    \end{equation}
    If the consistency of all $\mathrm{result}_{i,j,l}$ values is confirmed as in Eq.~(\ref{eq:55_57}), $\mathrm{CH}_{i,j}$ verifies the correctness of the aggregated signature as follows:
    \begin{equation}
    g^{\mathrm{H}(\mathrm{result}_{i,j,1})} \stackrel{?}{=} \mathrm{sig}_{\mathrm{CMs}} \cdot \mathrm{pk}_{\mathrm{CMs}},
    \label{eq:58}
    \end{equation}   
    where the aggregated signature $\mathrm{sig}_{\mathrm{CMs}}$ and public key $\mathrm{pk}_{\mathrm{CMs}}$ are computed as: 
    \begin{equation}
    \left\{
    \begin{aligned}
    \mathrm{sig}_{\mathrm{CMs}} &= \prod_{l=1}^{N_{i,j,\mathrm{CM}}} \mathrm{sig}_{\mathrm{CM}_{i,j,l}}^{s_{i,j,l}}, \\
    \mathrm{pk}_{\mathrm{CMs}} &= \left( \prod_{l=1}^{N_{i,j,\mathrm{CM}}} \mathrm{pk}_{\mathrm{CM}_{i,j,l}} \right)^{M}.
    \end{aligned}
    \right.
    \label{eq:59_60}
    \end{equation}
    If Eq.~(\ref{eq:58}) holds, then $\mathrm{CH}_{i,j}$ broadcasts the message $\{ \mathrm{sig}_{\mathrm{CMs}}, \mathrm{pk}_{\mathrm{CMs}}, c_{\mathrm{CH}_{i,j}}, Q_{\mathrm{CH}_{i,j}}, \mathrm{T}_{2} \}$ to each neighboring $\mathrm{CH}_{i,n}$, where $n \in [1, N_{i,\mathrm{CH}}], n \ne j$, and $\mathrm{T}_{2}$ is a newly generated timestamp. $c_{\mathrm{CH}_{i,j}}$ and  $Q_{\mathrm{CH}_{i,j}}$ are computed as follows:
    \begin{equation}
    \left\{
    \begin{aligned}
    c_{\mathrm{CH}_{i,j}} &= \mathrm{result}_{i,j,1} \oplus \mathrm{CT} \oplus \mathrm{T}_{2}, \\
    Q_{\mathrm{CH}_{i,j}} &= \mathrm{H}(\mathrm{result}_{i,j,1}, \mathrm{T}_{2}).
    \end{aligned}
    \right.
    \label{eq:61_62}
    \end{equation}

    \item Upon receiving the message from $\mathrm{CH}_{i,j}$, each neighboring $\mathrm{CH}_{i,n}$, where $n \in [1, N_{i,\mathrm{CH}}], n \ne j$, first verifies the freshness of timestamp $\mathrm{T}_{2}$. If $\mathrm{T}_{2}$ is valid, $\mathrm{CH}_{i,n}$ proceeds to compute the following equation:
    \begin{equation}
    \mathrm{result}_{i,j,1}' = c_{\mathrm{CH}_{i,j}} \oplus \mathrm{CT} \oplus \mathrm{T}_{2}.
    \end{equation}
    Then $\mathrm{CH}_{i,n}$ verifies the following equation:
    \begin{equation}
    Q_{\mathrm{CH}_{i,j}} \stackrel{?}{=} \mathrm{H}(\mathrm{result}_{i,j,1}', \mathrm{T}_{2}).
    \label{eq:63_64}
    \end{equation}
    If Eq.~(\ref{eq:63_64}) holds, $\mathrm{CH}_{i,n}$ continues by verifying the integrity of the received authentication information through the following equation:
    \begin{equation}
    g^{\mathrm{result}_{i,j,1}'} \stackrel{?}{=} \mathrm{sig}_{\mathrm{CMs}} \cdot \mathrm{pk}_{\mathrm{CMs}}.
    \label{eq:65}
    \end{equation}
    If Eq.~(\ref{eq:65}) is satisfied, then $\mathrm{CH}_{i,n}$ generates a confirmation message and replies to $\mathrm{CH}_{i,j}$ with $\{ Q_{\mathrm{CH}_{i,n}}, \mathrm{T}_{2} \}$, where $Q_{\mathrm{CH}_{i,n}}$ is computed as follows:
    \begin{equation}
    Q_{\mathrm{CH}_{i,n}} = \mathrm{H}(\mathrm{result}_{i,j,1}', \mathrm{T}_{2}).
    \label{eq:66}
    \end{equation}

    \item Upon receiving responses from the set of $\{ \mathrm{CH}_{i,n} \}_{n=1, n \ne j}^{N_{i,\mathrm{CH}}}$, $\mathrm{CH}_{i,j}$ verifies the freshness of timestamp $\mathrm{T}_{2}$. If valid, it proceeds to check whether the following condition holds:
    \begin{equation}
    Q_{\mathrm{CH}_{i,n}} \stackrel{?}{=} \mathrm{H}(\mathrm{result}_{i,j,1}, \mathrm{T}_{2}).
    \label{eq:67}
    \end{equation}
    If Eq.~\eqref{eq:67} is satisfied for all $\mathrm{CH}_{i,n}$, then $\mathrm{CH}_{i,j}$ sends the set of identifiers $\{ \mathrm{PID}_{\mathrm{NUAV}_{k}} \}_{k=1}^{N_{\mathrm{NUAV}}}$ to the corresponding ground base station $\mathrm{GBS}_{i}$. Upon receiving the identifiers, $\mathrm{GBS}_{i}$ stores them in its local database for subsequent management. Following the storage confirmation, $\mathrm{CH}_{i,j}$ sends the message $\{ \mathrm{res}_{k}, \mathrm{pk}_{\mathrm{CH}_{i,j}} \}$ to each $\mathrm{NUAV}_{k}$, where the response value $\mathrm{res}_{k}$ is computed as follows:
    \begin{equation}
    \mathrm{res}_{k} = \mathrm{H}(\mathrm{H}(\mathrm{CJT}_{i,j}), \mathrm{PID}_{\mathrm{NUAV}_{k}}, \mathrm{pk}_{\mathrm{CH}_{i,j}}).
    \label{eq:68}
    \end{equation}
    
    \item Upon receiving the message from $\mathrm{CH}_{i,j}$, each $\mathrm{NUAV}_{k}$ verifies the authenticity of $\mathrm{CH}_{i,j}$ by locally computing the following hash value:
    \begin{equation}
    \mathrm{res}_{k}' = \mathrm{H}(\mathrm{H}(\mathrm{CJT}_{i,j}), \mathrm{PID}_{\mathrm{NUAV}_{k}}).
    \label{eq:69}
    \end{equation}
    If $\mathrm{res}_{k}'$ matches the received $\mathrm{res}_{k}$, i.e., $\mathrm{res}_{k}' = \mathrm{res}_{k}$, then $\mathrm{NUAV}_{k}$ successfully authenticates $\mathrm{CH}_{i,j}$ and completes the cluster authentication process.

\end{enumerate}

Consequently, all $\mathrm{NUAV}_{k}$ ($k \in [1, N_{\mathrm{NUAV}}]$) are successfully authenticated and integrated into the $j$-th UAV cluster under the control of $\mathrm{GBS}_{i}$.

\subsubsection{Cross-cluster Phase with \textbf{LC2Am}}

When a UAV cluster requires additional UAVs to perform its designated tasks, its CH can request support from other CHs by recruiting UAVs from their clusters. These recruited UAVs are referred to as EUAVs. Prior to their integration, the requesting CH must authenticate the EUAVs, a process known as cross-cluster authentication, which is implemented via \textbf{LC2Am}.

\textbf{Cross-cluster Authentication:} Let $\mathrm{EUAV}_{i,j,l}$ denote a UAV originating from the $j$-th cluster of $\mathrm{GBS}_{i}$, with $\mathrm{CH}_{i,j}$ representing its CH. Let $\mathrm{CH}_{l,n}$ be the CH of the destination cluster, where $l \in [1, N_{\mathrm{GBS}}]$ and $n \in [1, N_{l,\mathrm{CH}}]$. Note that $\mathrm{CH}_{i,j}$ and $\mathrm{CH}_{l,n}$ belong to different clusters, which may or may not be under the administration of the same $\mathrm{GBS}$.The detailed steps are as follows:

\begin{enumerate}[label=\textbf{\textit{Step \arabic*: }}, align=left, leftmargin=0pt, labelsep=0em]

    \item $\mathrm{CH}_{i,j}$ sends a message $\{ C_{i,j}, \mathrm{PID}_{\mathrm{EUAV}_{i,j,l}}, \mathrm{T}_{3} \}$ to $\mathrm{CH}_{l,n}$, where $\mathrm{T}_{3}$ denotes a timestamp and $C_{i,j}$ is computed as follows:
    \begin{equation}
        C_{i,j} = \mathrm{H}(\mathrm{PID}_{\mathrm{EUAV}_{i,j,l}}, \mathrm{T}_{3}, \mathrm{CT}) \oplus \mathrm{CT}.
        \label{eq:70}
    \end{equation}

    \item Upon receiving the message, $\mathrm{CH}_{l,n}$ first checks the freshness of $\mathrm{T}_{3}$. If valid, it verifies the following equation:
    \begin{equation}
        \mathrm{H}(\mathrm{PID}_{\mathrm{EUAV}_{i,j,l}}, \mathrm{T}_{3}, \mathrm{CT}) \stackrel{?}{=} C_{i,j} \oplus \mathrm{CT}.
        \label{eq:71}
    \end{equation}
    If the verification succeeds, $\mathrm{CH}_{l,n}$ queries $\mathrm{GBS}_{l}$ to confirm whether $\mathrm{PID}_{\mathrm{EUAV}_{i,j,l}}$ exists in its database. If a match is found, $\mathrm{CH}_{l,n}$ accepts $\mathrm{EUAV}_{i,j,l}$ and assigns it a new pseudonymous identity, computed as follows:
    \begin{equation}
        \mathrm{PID}_{\mathrm{EUAV}_{i,j,l}}^{\mathrm{new}} = \mathrm{H}(\mathrm{PID}_{\mathrm{EUAV}_{i,j,l}}, \mathrm{T}_{3}, \mathrm{CT}).
        \label{eq:72}
    \end{equation}
    Subsequently, $\mathrm{CH}_{l,n}$ sends $\mathrm{PID}_{\mathrm{EUAV}_{i,j,l}}^{\mathrm{new}}$ to $\mathrm{GBS}_{l}$, which updates its internal database accordingly.

\end{enumerate}

\subsubsection{Cluster Session Key Update Phase with \textbf{CSKUm}}

When UAVs join or leave a cluster, the corresponding cluster session key must be updated to preserve forward and backward secrecy. This update process is realized by the proposed \textbf{CSKUm}.

\textbf{Cluster Session Key Update:} Let $\mathrm{CH}_{i,j}$ represent the CH of the $j$-th cluster managed by $\mathrm{GBS}_{i}$, and $\{ \mathrm{CM}_{i,j,l} \}_{l=1}^{N_{i,j,\mathrm{CM}}}$ denote the associated CMs. Let $\mathrm{key}_{i,j}$ be the previous session key. The detailed steps are described as follows:

\begin{enumerate}[label=\textbf{\textit{Step \arabic*: }}, align=left, leftmargin=0pt, labelsep=0em]

    \item $\mathrm{CH}_{i,j}$ generates a new session key $\mathrm{key}_{i,j}^{\mathrm{new}} \leftarrow \mathbb{Z}_{p}^{*}$, a fresh timestamp $\mathrm{T}_{4}$, and constructs a random polynomial of degree $N_{i,j,\mathrm{CM}} - 1$:
    \begin{equation}
        f(x) = \mathrm{key}_{i,j}^{\mathrm{new}} + b_{1}x + \cdots + b_{N_{i,j,\mathrm{CM}}-1}x^{N_{i,j,\mathrm{CM}}-1}.
        \label{eq:73}
    \end{equation}
    Then, $\mathrm{CH}_{i,j}$ sends to each $\mathrm{CM}_{i,j,l}$ the message $\{\mathrm{T}_{4},$ $ F_{\mathrm{CM}_{i,j,l}}, \{ g^{f(x_{i,j,n})} \}_{n=1, n \ne l}^{N_{i,j,\mathrm{CM}}}, \mathrm{H}(\mathrm{key}_{i,j}^{\mathrm{new}}, \mathrm{T}_{4})\}$, where $\mathrm{T}_{4}$ is a timestamp and $x_{i,j,n}$ and $F_{\mathrm{CM}_{i,j,l}}$ are computed as follows:
    \begin{equation}
    \left\{
    \begin{aligned}
        x_{i,j,n} &= \mathrm{H}(\mathrm{PID}_{\mathrm{CM}_{i,j,n}}), \\
        F_{\mathrm{CM}_{i,j,l}} &= f\left( \mathrm{H}(\mathrm{PID}_{\mathrm{CM}_{i,j,l}}) \right) \oplus \mathrm{H}(\mathrm{sk}_{\mathrm{CM}_{i,j,l}}, \mathrm{T}_{4}).
    \end{aligned}
    \right.
    \label{eq:75}
    \end{equation}

    \item Upon receiving the message, each $\mathrm{CM}_{i,j,l}$ first checks the freshness of $\mathrm{T}_{4}$. If valid, it recovers its own share as follows:
    \begin{equation}
        f(x_{i,j,l})' = F_{\mathrm{CM}_{i,j,l}} \oplus \mathrm{H}(\mathrm{sk}_{\mathrm{CM}_{i,j,l}}, \mathrm{T}_{4}).
        \label{eq:76}
    \end{equation}
    Then, each $\mathrm{CM}_{i,j,l}$ sends the message $\{\{ U_{i,j,n} \}_{n=1, n \ne l}^{N_{i,j,\mathrm{CM}}}\}$ to all other CMs, where $\{ U_{i,j,n} \}_{n=1, n \ne l}^{N_{i,j,\mathrm{CM}}}$ is computed as follows:
    \begin{equation}
    \left\{ U_{i,j,n} = g^{f(x_{i,j,n}) f(x_{i,j,l})'} \oplus f(x_{i,j,l})' \right\}_{n=1, n \ne l}^{N_{i,j,\mathrm{CM}}}.
    \end{equation}

    \item After receiving responses from other members, each $\mathrm{CM}_{i,j,n}$ reconstructs the updated session key. First, it computes the following equation:
    \begin{equation}
        \left\{ \mathrm{key}_{i,j,l}^{\mathrm{new}} = U_{i,j,l} \oplus g^{f(x_{i,j,l})' f(x_{i,j,n})} \right\}_{l=1, l \ne n}^{N_{i,j,\mathrm{CM}}},
        \label{eq:78}
    \end{equation}
    followed by
    \begin{equation}
    \begin{aligned}
        & \mathrm{key}_{i,j}^{\mathrm{new}'} = \\
        & \sum_{l=1}^{N_{i,j,\mathrm{CM}}} \left( \mathrm{key}_{i,j,l}^{\mathrm{new}} \prod_{n=1, n \ne l}^{N_{i,j,\mathrm{CM}}} \left( - \frac{x_{i,j,n}}{x_{i,j,l} - x_{i,j,n}} \right) \right).
        \label{eq:79}
    \end{aligned}
    \end{equation}
    Finally, $\mathrm{CM}_{i,j,n}$ verifies the following equation:
    \begin{equation}
        \mathrm{H}(\mathrm{key}_{i,j}^{\mathrm{new}'}, \mathrm{T}_{4}) \stackrel{?}{=} \mathrm{H}(\mathrm{key}_{i,j}^{\mathrm{new}}, \mathrm{T}_{4}).
        \label{eq:77}
    \end{equation}
    If the verification holds, $\mathrm{key}_{i,j}^{\mathrm{new}'}$ is accepted as the valid updated cluster session key.

\end{enumerate}

\section{Security Analysis}

This section provides the security analysis for security goals \textbf{S1}--\textbf{S7} introduced in Section~3.2.3. Specifically, \textbf{S1}--\textbf{S5} are analyzed by illustrative analysis while \textbf{S6} and \textbf{S7} are analyzed by formal analysis. Furthermore, the security goal comparison between related works and LP2-CASKU is presented to highlight the superiority of LP2-CASKU.

\subsection{Formal Analysis}

We conduct formal analysis for the security goals \textbf{S6} and \textbf{S7} by proving the following \textit{\textbf{Theorems~1}} and\textit{\textbf{~2}}, respectively.

\textit{\textbf{Theorem 1:}}  
For any PPT adversary $\mathcal{A}_{1}$, if Eq.~(\ref{eq:80}) holds under the random oracle model:
\begin{equation}
Adv_{\mathcal{A}_{1}}^{\mathrm{LP2\text{-}CASKU}} = \left| \Pr \left( \mathrm{win}_{\mathcal{A}_{1}}^{\mathrm{DUG}} \right) \right| < \varepsilon,
\label{eq:80}
\end{equation}
where $Adv_{\mathcal{A}_{1}}^{\mathrm{LP2\text{-}CASKU}}$ is the advantage of $\mathcal{A}_{1}$ in winning DUG defined in Section~3.2.4 and $\Pr \left( \mathrm{win}_{\mathcal{A}_{1}}^{\mathrm{DUG}} \right)$ denotes the success probability of $\mathcal{A}_{1}$, then LP2-CASKU satisfies the security goal \textbf{S6} defined in Section~3.2.3. Here,  $\varepsilon$ denotes a negligible function in the security parameter $\lambda$ of LP2-CASKU, meaning that for any positive polynomial $q(\cdot)$, there exists some point beyond as follows:
\begin{equation}
|\varepsilon(\lambda)| < \frac{1}{q(\lambda)}.
\end{equation}
Thus, $\mathcal{A}_{1}$'s advantage in winning the DUG is so small that it can be considered asymptotically equivalent to zero. In cryptographic terms, a scheme is deemed secure if no probabilistic polynomial-time adversary achieves more than a negligible advantage~\cite{goldreich2001foundations,katz2007introduction}.

\vspace{0.5em}

\textit{\textbf{Theorem 2:}}  
For any PPT adversary $\mathcal{A}_{2}$, if Eq.~(\ref{eq:81}) holds under the random oracle model:
\begin{equation}
Adv_{\mathcal{A}_{2}}^{\mathrm{LP2\text{-}CASKU}} = \left| \Pr \left( \mathrm{win}_{\mathcal{A}_{2}}^{\mathrm{DCG}} \right) - \frac{1}{2} \right| < \varepsilon,
\label{eq:81}
\end{equation}
where $Adv_{\mathcal{A}_{2}}^{\mathrm{LP2\text{-}CASKU}}$ is the advantage of $\mathcal{A}_{2}$ in winning DCG defined in Section~3.2.4, $\Pr \left( \mathrm{win}_{\mathcal{A}_{2}}^{\mathrm{DCG}} \right)$ denotes the success probability of $\mathcal{A}_{2}$, then LP2-CASKU satisfies the security goal \textbf{S7} defined in Section~3.2.3. $\varepsilon$ is as defined in \textit{\textbf{Theorem~1}}.

Due to the space limitation, we present the proof of \textbf{\textit{Theorems~1}} and\textbf{\textit{~2}} in the supplementary file.

\subsection{Illustrative Analysis}

In this section, we present the illustrative analysis about the security goals \textbf{S1}--\textbf{S5} introduced in Section~3.2.3. The detail is as follows:
\begin{itemize}

\item \textbf{S1) Authenticity of NUAVs and EUAVs:}  
The proposed scheme ensures the authenticity of both NUAVs and EUAVs. For NUAVs, authenticity is verified in two stages. First, each CM verifies the aggregated signature and ciphertext $\{\mathrm{sig}_{\mathrm{NUAVs}}, c_{\mathrm{NUAVs}}\}$ via Eq.~(\ref{eq:48}), which are derived from $\{ \mathrm{sig}_{k} \}_{k=1}^{N_{\mathrm{NUAV}}}$ and $\{ c_{k} \}_{k=1}^{N_{\mathrm{NUAV}}}$. Since generating a valid $\mathrm{sig}_{k}$ requires knowledge of $\mathrm{H}(\mathrm{CJT}_{i,j})$, only legitimate NUAVs can produce valid authentication tokens. Subsequently, the CH further authenticates NUAVs by verifying Eq.~(\ref{eq:65}). For EUAVs, authenticity is achieved through cross-cluster authentication, where the receiving CH verifies Eq.~(\ref{eq:71}) and checks the existence of $\mathrm{PID}_{\mathrm{EUAV}_{i,j,l}}$ in the GBS database. As only legitimate EUAVs can obtain $C_{i,j}$ from their original cluster, and invalid identities are not recorded in the database, the scheme effectively mitigates \textbf{entity impersonation attacks}.

\vspace{0.5em}

\item \textbf{S2) Anonymity:}  
Each UAV in the system communicates using a pseudonymous identity rather than its true identity. This design prevents adversaries from mapping communication sessions to specific UAVs, thereby thwarting \textbf{entity identity inference attacks} and ensuring anonymity.

\vspace{0.5em}

\item \textbf{S3) Unlinkability:}  
During cross-cluster authentication, the transmitted pair $\{ C_{i,j}, \mathrm{PID}_{\mathrm{EUAV}_{i,j,l}} \}$ is updated upon successful authentication by assigning a new pseudonymous identity as shown in Eq.~(\ref{eq:72}). The one-way nature of Eq.~(\ref{eq:70}) ensures that an adversary cannot correlate multiple authentication sessions to the same EUAV, thereby achieving unlinkability and defending against \textbf{EUAV movement inference attacks}. This property is guaranteed by the proposed \textbf{LC2Am}.

\vspace{0.5em}

\item \textbf{S4) Forward secrecy:}  
To prevent newly NUAVs from inferring previous cluster session keys, the proposed \textbf{CSKUm} updates the cluster session key after each UAV joins. The updated cluster session key, derived via polynomial-based secret sharing and reconstruction (Eq.~(\ref{eq:79})), ensures that knowledge of the new cluster session key does not reveal any information about the previous cluster session key, thus preserving forward secrecy against \textbf{cluster session key inference attacks}.

\vspace{0.5em}

\item \textbf{S5) Backward secrecy:}  
Similarly, when a UAV leaves the cluster, the cluster session key is re-established using the same \textbf{CSKUm} procedure. Since the leaving UAV lacks access to the newly generated shares, it is unable to compute the updated cluster session key (Eq.~(\ref{eq:79})). This guarantees backward secrecy and protects against \textbf{cluster session key inference attacks} from departed or compromised UAVs.

\end{itemize}

\subsection{Comparison of Security Goals}

We present the comparison of security goals between LP2-CASKU and related works~\cite{ref13}--\cite{ref20} in Table~\ref{table3}. The security goals \textbf{S1}--\textbf{S5} are introduced in Section~3.2.3.

\begin{table}[!t]
\renewcommand{\arraystretch}{1.2}
\caption{Comparison of Security Goals}
\label{table3}
\centering
\begin{tabular}{lccccc}
\hline
\textbf{Ref.} & \textbf{S1} & \textbf{S2} & \textbf{S3} & \textbf{S4} & \textbf{S5} \\
\hline
BDLA+~\cite{ref13} 2022 & $\times$ & $\checkmark$ & $\times$ & $\checkmark$ & $\checkmark$ \\
BCDA+~\cite{ref14} 2022 & $\checkmark$ & $\checkmark$ & $\times$ & $\times$ & $\times$ \\
TAGKA~\cite{ref15} 2023 & $\times$ & $\checkmark$ & $\times$ & $\checkmark$ & $\checkmark$ \\
SwarmAuth~\cite{ref16} 2024 & $\checkmark$ & $\checkmark$ & $\times$ & $\checkmark$ & $\times$ \\
BASUV~\cite{ref17} 2024 & $\times$ & $\checkmark$ & $\times$ & $\times$ & $\times$ \\
IOOSC-U2G~\cite{ref18} 2024 & $\times$ & $\checkmark$ & $\checkmark$ & $\checkmark$ & $\times$ \\
LBMA+~\cite{ref19} 2024 & $\times$ & $\checkmark$ & $\times$ & $\times$ & $\times$ \\
SAAF-IoD+~\cite{ref20} 2024 & $\times$ & $\checkmark$ & $\times$ & $\checkmark$ & $\times$ \\
LP2-CASKU ours 2025 & $\checkmark$ & $\checkmark$ & $\checkmark$ & $\checkmark$ & $\checkmark$ \\
\hline
\end{tabular}
\end{table}

\section{Performance Analysis}

In this section, we present a comprehensive performance evaluation of the proposed LP2-CASKU. First, we conduct a theoretical analysis of LP2-CASKU in terms of both computation and communication overheads. We further compare these overheads against those of representative baseline schemes to highlight the efficiency of LP2-CASKU. In addition, we implement simulation experiments using the Omnet++ framework~\cite{ref21} to assess the practical performance of LP2-CASKU and to validate the effectiveness of the proposed \textbf{MAm} in reducing communication overhead, which is critical for maintaining low latency in low-altitude economy networks environments. Finally, we evaluate the extent to which LP2-CASKU fulfills the performance goals \textbf{P1} and \textbf{P2} as defined in Section~3.2.3.

\subsection{Computation Overhead Analysis}

\begin{table}[!t]
\renewcommand{\arraystretch}{1.2}
\caption{Notations of Basic Operation Time Cost}
\label{table4}
\centering
\begin{tabular}{ll}
\hline
\textbf{Notation} & \textbf{Operations} \\
\hline
$T_{\mathrm{HF}}$ & \textrm{Hash function} \\
$T_{\mathrm{ME}}$ & \textrm{Modular exponential operation} \\
$T_{\mathrm{MM}}$ & \textrm{Modular multiplication operation} \\
$T_{\mathrm{PUF}}$ & \textrm{PUF operation} \\
$T_{\mathrm{XOR}}$ & \textrm{XOR operation} \\
$T_{\mathrm{ECPM}}$ & \textrm{Elliptic curve point multiplication operation} \\
$T_{\mathrm{ECPA}}$ & \textrm{Elliptic curve point addition operation} \\
$T_{\mathrm{BM}}$ & \textrm{Bilinear mapping operation} \\
$T_{\mathrm{SSS}}$ & \textrm{Secret shared shard operation} \\
$T_{\mathrm{FGA}}$ & \textrm{Fuzzy generator algorithm} \\
$T_{\mathrm{E}}$ & \textrm{Symmetric encryption} \\
$T_{\mathrm{D}}$ & \textrm{Symmetric decryption} \\
\hline
\end{tabular}
\end{table}

Given the heterogeneity in experimental settings, such as differences in hardware platforms and software environments, across existing schemes for evaluating computation overhead, we adopt a generalized and fair approach to represent the computation cost. Specifically, we quantify the overhead in terms of the number of fundamental cryptographic operations involved. Table~\ref{table4} summarizes the notations used to denote these basic operations for LP2-CASKU and the compared baseline schemes.

Using the notations from Table~\ref{table4}, we compile the computation overheads of LP2-CASKU and representative related works~\cite{ref13}--\cite{ref20} in Table~\ref{table5}, categorized by the number of operations incurred in each scheme. To ensure consistency in comparison, we divide the authentication workflows into three stages: (i) initialization, (ii) UAV authentication, and (iii) cluster session key update. The initialization stage corresponds to system bootstrapping and entity registration, while the UAV authentication stage includes intra-cluster joining and inter-cluster (cross-cluster) authentication procedures. The cluster session key update stage refers to the dynamic session key update process to ensure secure communication among authenticated UAVs.

In the context of LP2-CASKU, the initialization stage encompasses the Setup and Registration Phases as detailed in Sections~4.2.1 and 2. The UAV authentication stage covers the Join and Cross-cluster Authentication Phases described in Sections~4.2.3 and 4. The cluster session key update stage pertains to the process outlined in Section~4.2.5. As presented in Table~\ref{table5}, LP2-CASKU incurs substantially lower computation overhead compared to schemes such as~\cite{ref13,ref14,ref17,ref18}, which are based on elliptic curve cryptography. Notably, the scheme in~\cite{ref18} employs bilinear pairing operations, which are known to be computationly expensive, typically requiring a computation time approximately two orders of magnitude greater than modular multiplication operations (i.e., $10^2 \times$).

Unlike schemes~\cite{ref14}--\cite{ref16} and~\cite{ref20}, which involve symmetric encryption and decryption mechanisms, LP2-CASKU eliminates such reliance, thereby avoiding associated performance bottlenecks. Although LP2-CASKU exhibits a relatively larger number of basic operations, its overall computation overhead remains low due to the predominant use of lightweight primitives such as hash functions and XOR operations. Moreover, LP2-CASKU supports the concurrent authentication of multiple NUAVs, which introduces an additional overhead of $N_{\mathrm{NUAV}} T_{\mathrm{MM}}$, where $T_{\mathrm{MM}}$ denotes the cost of a modular multiplication. This multi-entity support is absent in most of the compared schemes, which are limited to single-entity authentication.

Furthermore, LP2-CASKU incorporates a dedicated cluster session key update mechanism(\textbf{CSKUm})—a critical feature for ensuring both forward and backward secrecy—which is not supported in~\cite{ref13}--\cite{ref15} and~\cite{ref17}--\cite{ref20}. Compared to~\cite{ref16}, LP2-CASKU incurs slightly higher basic operation counts but uniquely guarantees backward secrecy, which~\cite{ref16} does not address. It is worth emphasizing that the hash and XOR operations employed in LP2-CASKU are inherently efficient and impose negligible computation burden.

In conclusion, although LP2-CASKU introduces additional computation overhead to support cross-cluster authentication and secure key updates, its total computation cost remains acceptable. This efficiency is attributed to the use of lightweight cryptographic operations, making LP2-CASKU well-suited for deployment in computation-constrained environments, such as those found in low-altitude economy networks scenarios~\cite{ref28}.

\begin{table*}[htbp]
\scriptsize
\renewcommand{\arraystretch}{1.2}
\setlength{\tabcolsep}{2pt}
\centering
\begin{threeparttable}
\caption{Comparison of Computation Overhead Between LP2-CASKU and Related Works}
\label{table5}
\begin{tabular}{>{\raggedleft\arraybackslash}p{2.6 cm} >{\raggedright\arraybackslash}p{4.6 cm} >{\raggedright\arraybackslash}p{4.2 cm} >{\raggedright\arraybackslash}p{4.5 cm}}
\hline
\multicolumn{1}{c}{Ref.} & \multicolumn{1}{c}{Initialization} & \multicolumn{1}{c}{UAV Authentication} & \multicolumn{1}{c}{Cluster Session Key Update} \\
\hline
BDLA+~\cite{ref13}  2022 &
$2T_{\mathrm{HF}} + T_{\mathrm{XOR}} + 5T_{\mathrm{ECPM}}
$ &
$(2N_{\mathrm{CM}} + 6)T_{\mathrm{HF}} + T_{\mathrm{MM}} + 2T_{\mathrm{ECPM}} + 2T_{\mathrm{ECPA}}$ & Not applicable \\
\hline
BCDA+~\cite{ref14}  2022 &
$ 3N_{\mathrm{GBS}}T_{\mathrm{HF}} + 3N_{\mathrm{GBS}}T_{\mathrm{XOR}} + N_{\mathrm{GBS}}T_{\mathrm{ECPM}} $ $ ^a $ &
$4T_{\mathrm{HF}} + 18T_{\mathrm{ECPM}} + 10T_{\mathrm{ECPA}} + T_{\mathrm{BM}} + 3T_{\mathrm{E}} + 3T_{\mathrm{D}}
$ & Not applicable \\
\hline
TAGKA~\cite{ref15}  2023 &
${N_{\mathrm{CM}}T_{\mathrm{SSS}}}^{b}$ &
$13T_{\mathrm{HF}} + (2N_{\mathrm{CM}}^2 + N_{\mathrm{CM}})T_{\mathrm{MM}} + (3N_{\mathrm{CM}}^2 + 2)T_{\mathrm{XOR}} + 2T_{\mathrm{E}} + 2T_{\mathrm{D}}$ & Not applicable \\
\hline
SwarmAuth~\cite{ref16} 2024 &
$2T_{\mathrm{HF}} + T_{\mathrm{PUF}} + 11T_{\mathrm{XOR}}
$ &
$9T_{\mathrm{HF}} + T_{\mathrm{PUF}} + (2N_{\mathrm{CM}} + 14)T_{\mathrm{XOR}} + 2N_{\mathrm{CM}}T_{\mathrm{E}}
$ &
$N_{\mathrm{CM}}^2 T_{\mathrm{MM}}
$ \\
\hline
BASUV~\cite{ref17}  2024 &
$T_{\mathrm{HF}} + (N_{\mathrm{GBS}} - 1)T_{\mathrm{MM}} + (N_{\mathrm{GBS}} + 1)T_{\mathrm{ECPM}} + N_{\mathrm{GBS}}T_{\mathrm{ECPA}} + T_{\mathrm{BM}}
$ &
$6T_{\mathrm{HF}} + 2T_{\mathrm{MM}} + 4T_{\mathrm{XOR}} + 2T_{\mathrm{ECPM}} + 4T_{\mathrm{BM}}
$ & Not applicable \\
\hline
IOOSC-U2G~\cite{ref18} 2024 &
$ T_{\mathrm{HF}} + T_{\mathrm{MM}} + 3T_{\mathrm{ECPM}}
$ &
$ 5T_{\mathrm{HF}} + 2T_{\mathrm{XOR}} + T_{\mathrm{MM}}  + 6T_{\mathrm{ECPM}} + T_{\mathrm{ECPA}}
$ & Not applicable \\
\hline
LBMA+~\cite{ref19}  2024 &
$ 6T_{\mathrm{HF}}$ &
$ 20T_{\mathrm{HF}} + 14T_{\mathrm{XOR}}
$ & Not applicable \\
\hline
SAAF-IoD+~\cite{ref20} 2024 &
$ 2T_{\mathrm{HF}} + T_{\mathrm{FGA}} + T_{\mathrm{E}}
$ &
$ 13T_{\mathrm{HF}} + 20T_{\mathrm{XOR}} + T_{\mathrm{FGA}} +  3T_{\mathrm{E}} + 3T_{\mathrm{D}}
$ & Not applicable \\
\hline
LP2-CASKU Ours 2025 &
$ 3T_{\mathrm{HF}} + 2T_{\mathrm{ME}} + T_{\mathrm{MM}}
$ &
$ (N_{\mathrm{CM}} + 18)T_{\mathrm{HF}} + 8T_{\mathrm{ME}} + (N_{\mathrm{NUAV}} + 2N_{\mathrm{CM}} + 3)T_{\mathrm{MM}} + (N_{\mathrm{CM}} + 10)T_{\mathrm{XOR}}
$ $^{c}$ &
$ (N_{\mathrm{CM}} + 2)T_{\mathrm{HF}} + (3N_{\mathrm{CM}} - 1)T_{\mathrm{ME}} + (N_{\mathrm{CM}}^2 - 1)T_{\mathrm{MM}} + (2N_{\mathrm{CM}} + 10)T_{\mathrm{XOR}} + N_{\mathrm{CM}}T_{\mathrm{SSS}}
$ \\
\hline
\end{tabular}

\begin{tablenotes}
\footnotesize
\item[a-c] $N_{\mathrm{GBS}}$ is the number of GBSs. \quad $N_{\mathrm{CM}}$ is the number of CMs. \quad $N_{\mathrm{NUAV}}$ is the number of NUAVs.
\end{tablenotes}
\end{threeparttable}
\vspace{-10pt}
\end{table*}

\subsection{Communication Overhead Analysis}

To ensure a consistent and fair comparison, we adopt a generalized approach to assess the communication overhead of LP2-CASKU and related schemes. Specifically, we quantify communication costs in terms of the number and size of basic elements transmitted during protocol execution. The notations representing the lengths of basic elements are summarized in Table~\ref{table6}. Based on these notations, Table~\ref{table7} presents the communication overheads of LP2-CASKU and representative schemes~\cite{ref13}--\cite{ref20} across three functional stages, consistent with those defined in Table~\ref{table5}.

As illustrated in Table~\ref{table7}, LP2-CASKU primarily utilizes lightweight cryptographic primitives, with hash function outputs represented as elements in $\mathbb{Z}_{p}^{*}$. Consequently, the protocol exchanges are composed mainly of group elements from $\mathbb{Z}_{p}$ and timestamps, similar to the design in~\cite{ref19}. To support dynamic cluster scalability, LP2-CASKU introduces collaborative authentication mechanisms involving multiple CHs. This leads to an additional communication overhead of $4N_{\mathrm{CH}}|\mathbb{Z}_{p}| + N_{\mathrm{CH}}|\mathrm{T}|$ bits for NUAV authentication, where $|\mathbb{Z}_{p}|$ and $|\mathrm{T}|$ denote the bit-lengths of group elements and timestamps, respectively.

Furthermore, the integration of the proposed \textbf{LC2Am} for cross-cluster authentication introduces additional communication costs. However, such operations are event-driven and occur only when EUAV mobility is needed. Although the communication overhead incurred during the Cluster Session Key Update Phase is higher than that of~\cite{ref16}, these updates are performed infrequently and remain within acceptable bounds for real-world deployments.

In addition, LP2-CASKU incorporates \textbf{MAm} to mitigate redundant transmissions during NUAV authentication. This optimization yields a communication reduction of $(5N_{\mathrm{NUAV}} + 2N_{\mathrm{CH}})|\mathbb{Z}_{p}^{*}|$ bits when authenticating $N_{\mathrm{NUAV}}$ NUAVs, thereby substantially improving transmission efficiency. This feature is particularly advantageous in bandwidth-constrained scenarios typical of low-altitude economy networks environments~\cite{ref29}.

In conclusion, LP2-CASKU achieves low overall communication overhead by leveraging compact cryptographic primitives. While collaborative authentication and cluster session key update procedures introduce additional communication costs, their infrequent occurrence ensures these costs remain acceptable. Meanwhile, the deployment of \textbf{MAm} significantly reduces overhead during NUAV onboarding, underscoring LP2-CASKU’s practicality and efficiency for low-bandwidth UAV networks.

\begin{table}[!t]
\renewcommand{\arraystretch}{1.2}
\caption{Notations of Basic Element Size}
\label{table6}
\centering
\begin{tabular}{ll}
\hline
\textbf{Notation} & \textbf{Element} \\
\hline
$|\mathrm{PUF()}|$ & Size of PUF output \\
$|\mathbb{G}_{E}|$ & Size of element in elliptic curve cyclic group $\mathbb{G}_{E}$ \\
$|\mathbb{Z}_{p}^{*}|$ & Size of element in integrity group $\mathbb{Z}_{p}^{*}$ \\
$|m|$ & Size of message $m$ \\
$|\mathrm{T}|$ & Size of timestamp $\mathrm{T}$ \\
$|V|$ & Row vector of matrix $M$ \\
$|CP|$ & Size of Chebyshev polynomial \\
\hline
\end{tabular}
\end{table}

\begin{table*}[htbp]
\scriptsize
\renewcommand{\arraystretch}{1.2}
\setlength{\tabcolsep}{2pt}
\centering
\begin{threeparttable}
\caption{Comparison of Communication Overhead Between LP2-CASKU and Related Works}
\label{table7}
\begin{tabular}{>{\raggedleft\arraybackslash}p{2.6 cm} p{4.6 cm} p{4.2 cm} p{4.5 cm}}
\hline
\multicolumn{1}{c}{Ref.} & \multicolumn{1}{c}{Initialization} & \multicolumn{1}{c}{UAV Authentication} & \multicolumn{1}{c}{Cluster Session Key Update} \\
\hline
BDLA+~\cite{ref13} 2022 &
$2|\mathbb{Z}_{p}^{*}|$ &
$|\mathbb{G}_{\mathrm{E}}| + (N_{\mathrm{M}} + 4)|\mathbb{Z}_{p}^{*}| + (N_{\mathrm{M}} + 1)|m| + |\mathrm{T}|$ & Not applicable \\
\hline
BCDA+~\cite{ref14} 2022 &
$(3N_{\mathrm{GBS}} + 1)|\mathbb{G}_{\mathrm{E}}| + (5N_{\mathrm{GBS}} + 7)|\mathbb{Z}_{p}^{*}| + 4|\mathrm{T}|$\tnote{a} &
$2N_{\mathrm{CM}}|\mathbb{G}_{\mathrm{E}}| + (7N_{\mathrm{CM}} - 1)|\mathbb{Z}_{p}^{*}| + (3N_{\mathrm{CM}} - 1)|\mathrm{T}|$\tnote{b} & Not applicable \\
\hline
TAGKA~\cite{ref15} 2023 &
$2N_{\mathrm{CM}}|\mathbb{Z}_{p}^{*}| + N_{\mathrm{CM}}|CP|$ &
$6(N_{\mathrm{CM}} - 1)|\mathbb{Z}_{p}^{*}| + 5(N_{\mathrm{CM}} - 1)|m| + 5(N_{\mathrm{CM}} - 1)|CP|$ & Not applicable \\
\hline
SwarmAuth~\cite{ref16} 2024 &
$6|\mathrm{PUF}()|$ &
$5|\mathrm{PUF}()| + 2(N_{\mathrm{CM}}^2 - N_{\mathrm{CM}})|m|$ &
 $4(N_{\mathrm{CM}} - 1)|\mathbb{Z}_{p}^{*}| + 2(N_{\mathrm{CM}} - 1)|m| + 2(N_{\mathrm{CM}} - 1)|CP|$ \\
\hline
BASUV~\cite{ref17} 2024 &
$|\mathbb{G}_{\mathrm{E}}| + 3|\mathbb{Z}_{p}^{*}|$ &
$3|\mathbb{G}_{\mathrm{E}}| + 13|\mathbb{Z}_{p}^{*}| + 2|m|$ & Not applicable \\
\hline
IOOSC-U2G~\cite{ref18} 2024 &
$2|\mathbb{G}_{\mathrm{E}}| + 2|\mathbb{Z}_{p}^{*}|$ &
$(2N_{\mathrm{CM}} + 3)|\mathbb{G}_{\mathrm{E}}| + (N_{\mathrm{CM}} + 4)|\mathbb{Z}_{p}^{*}| + (N_{\mathrm{CM}} + 1)|m| + 2|\mathrm{T}|$ & Not applicable \\
\hline
LBMA+~\cite{ref19} 2024 &
$(N_{\mathrm{CM}} + 8)|\mathbb{Z}_{p}^{*}| + |\mathrm{T}|$ &
$(5N_{\mathrm{CM}} + 3)|\mathbb{Z}_{p}^{*}| + (2N_{\mathrm{CM}} + 1)|\mathrm{T}|$ & Not applicable \\
\hline
SAAF-IoD+~\cite{ref20} 2024 &
$5|\mathbb{Z}_{p}^{*}| + 2|CP|$ &
$6|\mathbb{Z}_{p}^{*}| + 3|\mathrm{T}| + 3|CP|$ & Not applicable \\
\hline
LP2-CASKU Ours 2025 &
$10|\mathbb{Z}_{p}^{*}|$ &
$(6N_{\mathrm{CM}} + 4N_{\mathrm{CH}} + 18)|\mathbb{Z}_{p}^{*}| + (6N_{\mathrm{CM}} + N_{\mathrm{CH}} + 4)|\mathrm{T}|$\tnote{c} &
$(N_{\mathrm{CM}}^2 + 5N_{\mathrm{CM}} - 1)|\mathbb{Z}_{p}^{*}| + N_{\mathrm{CM}}|\mathrm{T}|$ \\
\hline
\end{tabular}

\begin{tablenotes}
\footnotesize
\item[a-c] $N_{\mathrm{GBS}}$ is the number of GBSs. \quad $N_{\mathrm{CM}}$ is the number of CMs. \quad $N_{\mathrm{NUAV}}$ is the number of NUAVs.
\end{tablenotes}
\end{threeparttable}
\end{table*}

\subsection{Simulation Experiments}

In this section, we conduct simulation-based experiments to evaluate the practical applicability of LP2-CASKU and to demonstrate the effectiveness of the proposed \textbf{MAm} in reducing communication overhead.

\begin{table}[!t]
\renewcommand{\arraystretch}{1.2}
\caption{The Configuration of Hardware and Software Platforms}
\label{table8}
\centering
\begin{tabular}{ll}
\hline
\textbf{Platform} & \textbf{Configuration} \\
\hline
CPU & Intel(R) Core(TM) i7-10700 CPU @ 2.90GHz \\
Operating System & Ubuntu 20.04 \\
Simulation Software & Omnet++ 6.0.3 / inet 4.5.4 \\
\hline
\end{tabular}
\end{table}

\subsubsection{Experiment Settings}

To evaluate the practicability of LP2-CASKU and validate the effectiveness of the proposed \textbf{MAm} in reducing communication overhead, we implement simulations using the Omnet++ network simulator in conjunction with the INET framework~\cite{ref30}. The hardware and software environments utilized for the simulation are summarized in Table~\ref{table8}.

To ensure the reliability and relevance of the simulation results, various practical considerations are incorporated into the experimental design. Key parameters related to network configuration, UAV mobility, and energy consumption are listed in Table~\ref{table9}. For a more comprehensive overview of simulation settings, please refer to the supplementary file.

\subsubsection{Latency Analysis}

To quantitatively evaluate the effectiveness of \textbf{MAm} in reducing overall latency through communication overhead mitigation, we compare the latency performance of LP2-CASKU under two configurations: with and without the use of \textbf{MAm}. When \textbf{MAm} is enabled, LP2-CASKU operates as described in Section~4, wherein the CH aggregates the join requests of NUAVs along with the authentication results provided by CMs to authenticate the NUAVs. In contrast, when \textbf{MAm} is disabled, the CH processes each NUAV’s join request and its corresponding authentication result from CMs independently, without aggregation.

It is evident that the latency of the join phase in LP2-CASKU is influenced by several key parameters, including the number of NUAVs ($N_{\mathrm{NUAV}}$), the number of CMs ($N_{\mathrm{CM}}$) within a cluster, and the number of CHs ($N_{\mathrm{CH}}$) in the UAV swarm. Accordingly, we perform a latency analysis under varying values of these parameters, which are modeled as discrete uniform distributions as follows:

\begin{itemize}
    \item $N_{\mathrm{NUAV}}$: Discrete uniform distribution over the set $\{3, 4, 5, 6, 7\}$.
    \item $N_{\mathrm{CM}}$: Discrete uniform distribution over the set $\{3, 4, 5, 6, 7\}$.
    \item $N_{\mathrm{CH}}$: Discrete uniform distribution over the set $\{3, 4, 5, 6, 7\}$.
\end{itemize}

We choose the discrete uniform distribution because, in the absence of precise real-world statistics, it represents an unbiased scenario in which each feasible value within the tested range is equally likely. This modeling assumption avoids introducing undue bias into simulation results and is widely used in network and performance evaluations when only bounded ranges are known~\cite{bogon2013distuniform, li2017swarmdensity}. It allows us to systematically explore system behavior across the entire plausible domain, ensuring that latency trends are not skewed toward any particular parameter value. Moreover, prior works in UAV cluster deployment have demonstrated that cluster sizes ranging from 7 to 15 UAVs are commonly used in both experimental and simulation studies~\cite{vasarhelyi2014flocking, kwa2023density}, validating the relevance of our selected range.

\begin{figure*}[htbp]
    \centering
    % 第一行三图
    \begin{minipage}[t]{0.32\textwidth}
        \includegraphics[width=\linewidth]{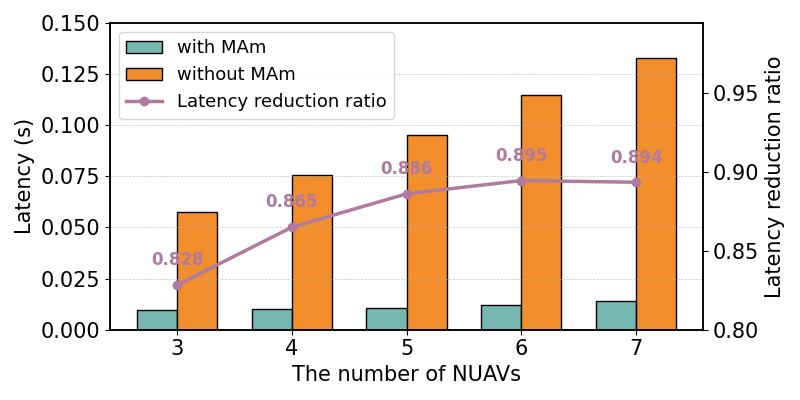}
        \subcaption*{(a) Latency of LP2-CASKU when $N_{\mathrm{NUAV}}$ grows and $N_{\mathrm{CM}} = N_{\mathrm{CH}} = 5$}
    \end{minipage}\hfill
    \begin{minipage}[t]{0.32\textwidth}
        \includegraphics[width=\linewidth]{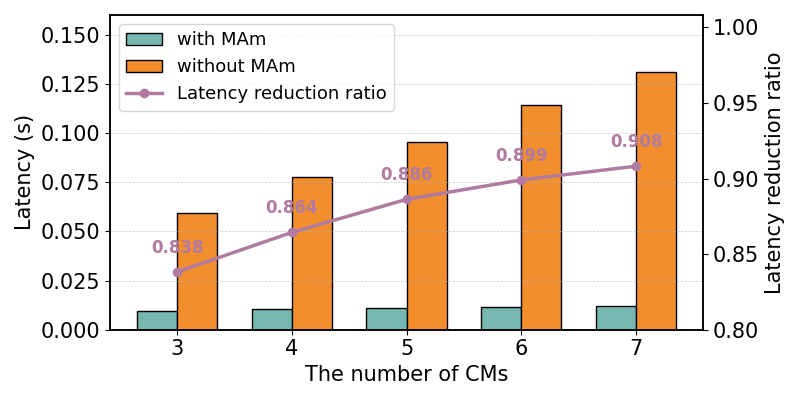}
        \subcaption*{(b) Latency of LP2-CASKU when $N_{\mathrm{CM}}$ grows and $N_{\mathrm{NUAV}} = N_{\mathrm{CH}} = 5$}
    \end{minipage}\hfill
    \begin{minipage}[t]{0.32\textwidth}
        \includegraphics[width=\linewidth]{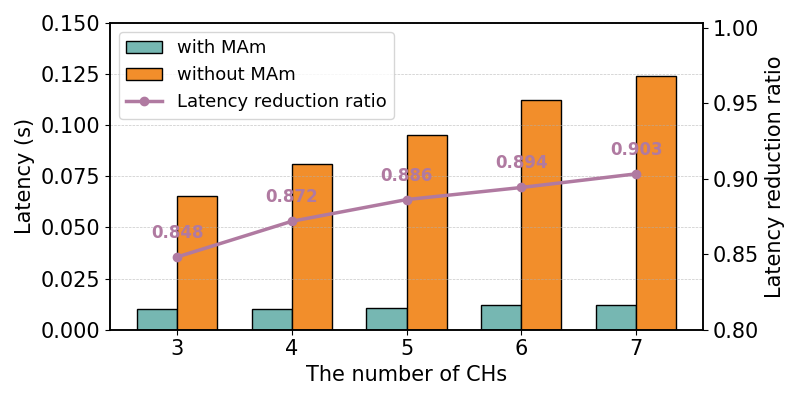}
        \subcaption*{(c) Latency of LP2-CASKU when $N_{\mathrm{CH}}$ grows and $N_{\mathrm{NUAV}} = N_{\mathrm{CM}} = 5$}
    \end{minipage}
\caption{Latency of LP2-CASKU in the Join Phase under varying $N_{\mathrm{NUAV}}$, $N_{\mathrm{CM}}$, and $N_{\mathrm{CH}}$, comparing scenarios with and without \textbf{MAm}. Results demonstrate that \textbf{MAm} effectively stabilizes latency as UAV swarm scale increases.}
\label{fig2}
\end{figure*}

The latency results are presented in Fig.~\ref{fig2}. As shown in Fig.~\ref{fig2}(a), when the number of NUAVs ($N_{\mathrm{NUAV}}$) increases from 3 to 7, the latency of LP2-CASKU with \textbf{MAm} exhibits only a modest increase from 9.86~ms to 14.15~ms. In contrast, the latency of LP2-CASKU without \textbf{MAm} grows substantially, from 57.43~ms to 132.98~ms. This indicates that \textbf{MAm} reduces the latency by approximately \textbf{82.8\%} to \textbf{89.5\%} as $N_{\mathrm{NUAV}}$ increases. The minor latency introduce by \textbf{MAm} is attributed to the aggregation overhead incurred during the join phase.

Consistent trends are observed in Fig.~\ref{fig2}(b) and Fig.~\ref{fig2}(c). As the number of CMs ($N_{\mathrm{CM}}$) and CHs ($N_{\mathrm{CH}}$) increases from 3 to 7, the latency of LP2-CASKU with \textbf{MAm} remains relatively stable, varying from 9.60~ms to 12.02~ms and from 9.95~ms to 12.00~ms, respectively. Conversely, without \textbf{MAm}, the latency increases significantly, reaching 130.92~ms and 124.09~ms, respectively. These results demonstrate that \textbf{MAm} reduces latency by approximately \textbf{83.8\%} to \textbf{90.8\%} under varying $N_{\mathrm{CM}}$, and \textbf{84.8\%} to \textbf{90.3\%} under varying $N_{\mathrm{CH}}$.

The observed latency reduction is primarily due to the ability of \textbf{MAm} to aggregate and broadcast authentication results from multiple CMs to other CHs in a single operation. This parallelized processing approach mitigates the impact of increasing cluster size on latency, thereby enhancing scalability and responsiveness in low-altitude UAV swarm networks.

\begin{figure}[!t]
\includegraphics[width=3.5in]{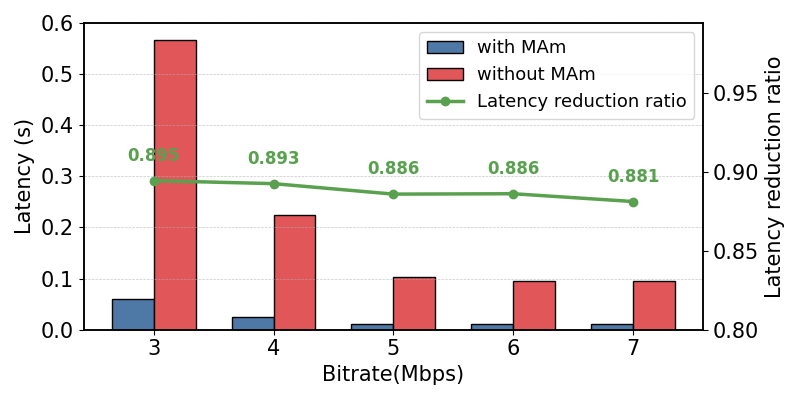}
\caption{Latency of LP2-CASKU under different network bitrates with and without \textbf{MAm}. Even at high bitrates, \textbf{MAm} significantly reduces latency by minimizing communication overhead.}
\label{fig3}
\vspace{-10pt}
\end{figure}

We further evaluate the latency performance of LP2-CASKU under varying network bitrates, with results presented in Fig.~\ref{fig3}. The quantitative analysis confirms the consistent effectiveness of \textbf{MAm} in reducing latency across diverse communication environments. Specifically, when the network bitrate is $1$ Mbps, LP2-CASKU with \textbf{MAm} achieves a latency of 59.72~ms, compared to 566.51~ms without \textbf{MAm}, yielding an \textbf{89.5\%} reduction. As the bitrate increases to $11$ Mbps, the latency with \textbf{MAm} decreases to 24.10~ms, whereas the baseline remains at 224.63~ms, reflecting an \textbf{89.3\%} reduction. Even under higher bitrates—$24$ Mbps, $48$ Mbps, and $54$ Mbps—\textbf{MAm} consistently maintains low latency ranging from 11.82~ms to 11.23~ms. In contrast, the corresponding baseline latency remains above 94.51~ms, indicating that \textbf{MAm} continues to reduce latency by over \textbf{88\%} in high-throughput conditions. These results demonstrate that \textbf{MAm} substantially mitigates communication overhead and ensures sustained low-latency performance, regardless of the underlying network bitrate.

In summary, the integration of \textbf{MAm} within LP2-CASKU leads to significant latency reduction, particularly under challenging network conditions such as limited bandwidth or large-scale UAV deployments. As illustrated in Fig.~\ref{fig2}, the latency remains stable despite increases in the number of NUAVs, CMs, and CHs, with reductions consistently exceeding \textbf{82.0\%}. Furthermore, Fig.~\ref{fig3} validates the resilience of LP2-CASKU under constrained bandwidth conditions, achieving an \textbf{89.5\%} latency reduction even at 1~Mbps.

These findings collectively confirm that LP2-CASKU effectively supports fast and scalable UAV cluster formation. In particular, its low-latency characteristics enable timely response and high service reliability, meeting the real-time communication demands inherent to low-altitude economy networks scenarios.

\begin{table}[!t]
\renewcommand{\arraystretch}{1.2}
\caption{The Parameter Settings of the Simulation Experiment}
\label{table9}
\centering
\begin{tabular}{p{0.8 cm} p{3.4 cm} p{3.4 cm}}
\hline
\textbf{Category} & \textbf{Name} & \textbf{Setting} \\
\hline
\multirow{6}{*}{Network} & Network configurator & Ipv4NetworkConfigurator \\
 & Arp mode & GlobalArp \\
 & Network interface & IEEE80211Interface \\
 & Management layer & IEEE80211MgmtAdhoc \\
 & Bitrate & 48 Mbps \\
 & Transport layer protocol & UDP \\
\hline
\multirow{6}{*}{UAV} & Physical layer modules & IEEE80211ScalarRadio \\
 & Transmitter power & 500 mW \\
 & Receiver sensitivity & -85 dBm \\
 & Receiver energy detection & -85 dBm \\
 & Mobility module & Linear movement \\
 & Mobility speed & 100 m/s \\
 & Constraint area & 2100 m $\times$ 2100 m \\
\hline
\multirow{7}{*}{Energy} & Power consumption model & StateBasedEpEnergyConsumer \\
 & Sleep power consumption & 0 mW \\
 & Receiver idle power consumption & 2 mW \\
 & Receiver busy power consumption & 5 mW \\
 & Receiver receiving power consumption & 100 mW \\
 & Transmitter idle power consumption & 2 mW \\
 & Transmitter transmitting power consumption & 300 mW \\
 & UAV initial energy & 0.01 J \\
\hline
\end{tabular}
\end{table}

\subsubsection{Energy Consumption Analysis}

\begin{figure}[!t]
\includegraphics[width=3.5in]{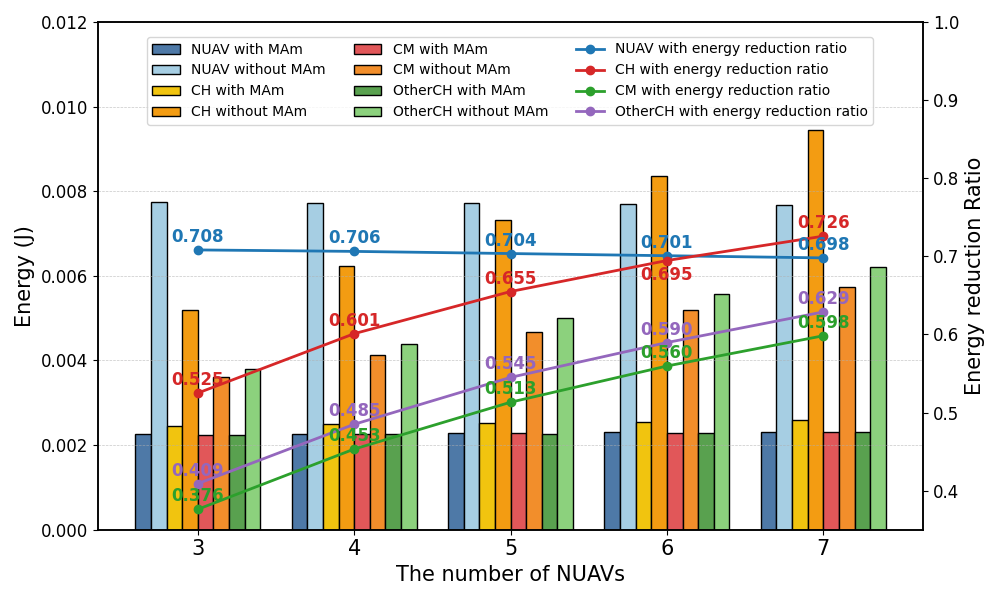}
\caption{Energy consumption of NUAV, CM, and CH under different numbers of NUAVs ($N_{\mathrm{NUAV}}$), with $N_{\mathrm{CM}} = N_{\mathrm{CH}} = 5$. \textbf{MAm} reduces overall energy consumption for CMs and CHs as $N_{\mathrm{NUAV}}$ grows.}
\label{fig4}
\vspace{-10pt}
\end{figure}

\begin{figure}[!t]
\includegraphics[width=3.5in]{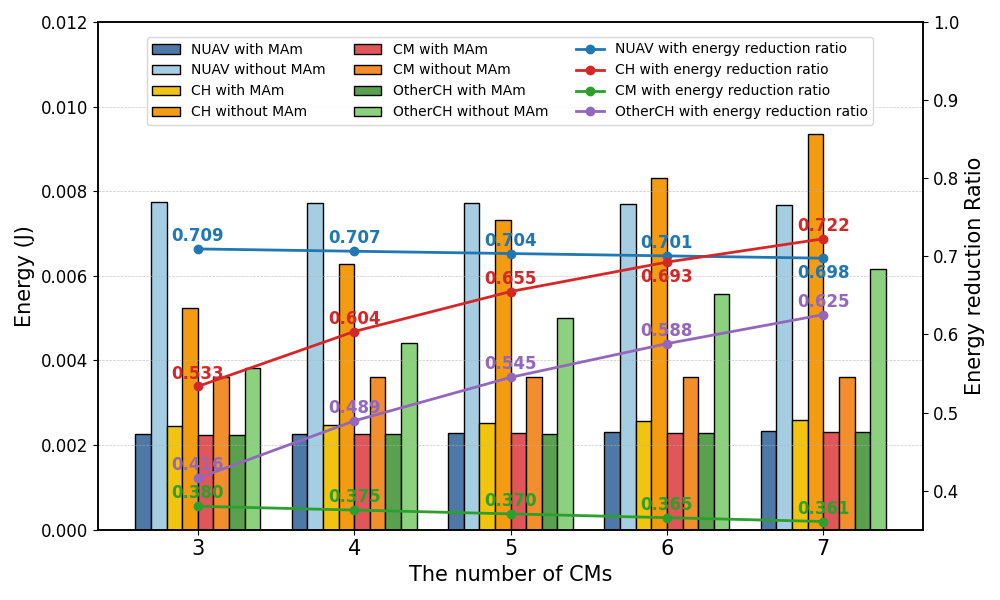}
\caption{Energy consumption of NUAV, CM, and CH under different numbers of CMs ($N_{\mathrm{CM}}$), with $N_{\mathrm{NUAV}} = N_{\mathrm{CH}} = 5$. \textbf{MAm} mitigates the increase in energy overhead for CHs and CMs as $N_{\mathrm{CM}}$ increases.}
\label{fig5}
\vspace{-10pt}
\end{figure}

\begin{figure}[!t]
\includegraphics[width=3.5in]{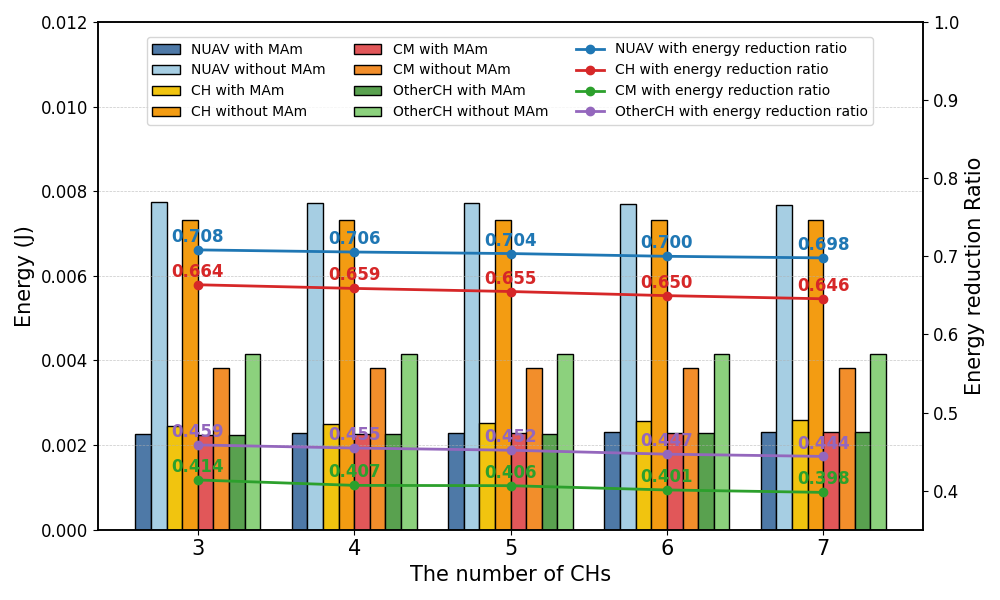}
\caption{Energy consumption of NUAV, CM, and CH under different numbers of CHs ($N_{\mathrm{CH}}$), with $N_{\mathrm{NUAV}} = N_{\mathrm{CM}} = 5$. The energy overhead of CMs and CHs remains stable, confirming the scalability of LP2-CASKU with \textbf{MAm}.}
\label{fig6}
\vspace{-10pt}
\end{figure}

To further evaluate the practicality of LP2-CASKU, we analyze its energy efficiency with and without the integration of \textbf{MAm}. Specifically, we investigate the energy consumption incurred by each NUAV, CM, and CH as the number of NUAVs, CMs, and CHs varies. The corresponding results are presented in Fig.~\ref{fig4} to Fig.~\ref{fig6}. Note that the CH in each figure refers to the head of the cluster that the NUAV joins, while the "other CHs" represent the CHs of remaining clusters within the UAV swarm.

Fig.~\ref{fig4} shows that as the number of NUAVs increases from 3 to 7, the energy consumption of both CHs and CMs in the baseline scheme (i.e., without \textbf{MAm}) is significantly higher compared to that under LP2-CASKU with \textbf{MAm}, which remains largely stable. Specifically, the energy consumption of the CH is reduced by approximately \textbf{72.6\%}, from $9.44 \times 10^{-3}$~J (without \textbf{MAm}) to $2.59 \times 10^{-3}$~J (with \textbf{MAm}). Similarly, the CM's energy consumption decreases by about \textbf{59.8\%}, from $5.75 \times 10^{-3}$~J to $2.25 \times 10^{-3}$~J. This reduction is primarily due to \textbf{MAm} enabling the CH to aggregate NUAV join requests and forward them in a batch to the CM, thereby eliminating redundant authentication operations. Moreover, the energy consumption of other CHs also drops by roughly \textbf{62.9\%}, from $6.22 \times 10^{-3}$~J to $2.25 \times 10^{-3}$~J, further confirming that the benefits of message aggregation extend across clusters.

As illustrated in Fig.~\ref{fig5}, increasing the number of CMs from 3 to 7 results in significantly higher energy consumption for both the CH and other CHs in the absence of \textbf{MAm}, while the energy usage with \textbf{MAm} remains nearly constant. Specifically, the CH’s energy consumption is reduced by approximately \textbf{72.2\%}, from $9.35 \times 10^{-3}$~J (without \textbf{MAm}) to $2.45 \times 10^{-3}$~J (with \textbf{MAm}). Similarly, the energy usage of other CHs decreases by around \textbf{62.5\%}, from $6.16 \times 10^{-3}$~J to $2.23 \times 10^{-3}$~J. These results highlight the efficiency of \textbf{MAm} in minimizing repeated authenticity verification transmissions, which would otherwise be performed independently by each CH for every CM.

As shown in Fig.~\ref{fig6}, variations in the number of CHs from 3 to 7 have minimal impact on the energy consumption of CMs, the designated CH, and other CHs, regardless of whether \textbf{MAm} is applied. Specifically, with \textbf{MAm}, the CH’s energy consumption remains nearly stable, ranging from $2.46 \times 10^{-3}$~J to $2.59 \times 10^{-3}$~J, while other CHs consistently consume approximately $2.25 \times 10^{-3}$~J. Similarly, the energy usage of CMs shows only slight variation, remaining below $2.27 \times 10^{-3}$~J. Even without \textbf{MAm}, the energy consumption does not increase significantly, as the introduction of additional CHs does not impose extra authentication or communication burdens on the existing CH or CMs. These results confirm the scalability of LP2-CASKU, demonstrating that it can efficiently support dynamic UAV networks without introducing additional energy overhead as the number of clusters grows.

Moreover, across all scenarios depicted in Fig.~\ref{fig4} to Fig.~\ref{fig6}, the energy consumption of NUAVs remains largely unaffected. Under \textbf{MAm}, NUAV energy usage fluctuates minimally between $2.25 \times 10^{-3}$~J and $2.32 \times 10^{-3}$~J, which is expected, as each NUAV independently transmits its join request regardless of system scale or configuration.

In conclusion, \textbf{MAm} introduces negligible computation overhead but yields substantial reductions in overall system energy consumption, particularly in large-scale deployments with numerous NUAVs and CMs. These findings validate the energy efficiency of LP2-CASKU and its suitability for resource-constrained environments, thereby aligning with the performance and cost-effectiveness requirements of low-altitude economy networks.

\subsection{Performance Goal Analysis}
This section analyzes the achievement of performance goals \textbf{P1} and \textbf{P2} as introduced in Section~3.2.3.

\begin{itemize}

\item \textbf{P1) Lightweight cross-cluster authentication:}  
In the proposed \textbf{LC2Am}, when a EUAV performs cross-cluster authentication, the incurred computation overhead is only $3T_{\mathrm{HF}} + 2T_{\mathrm{XOR}}$, and the corresponding communication overhead is $3|\mathbb{Z}_{p}^{*}| + |\mathrm{T}|$ bits. This is significantly lower than the overhead required for NUAV authentication, which involves $({N_{\mathrm{CM}}} + 15)T_{\mathrm{HF}} + 8T_{\mathrm{ME}} + (2N_{\mathrm{CM}} + 3)T_{\mathrm{MM}} + (N_{\mathrm{CM}} + 8)T_{\mathrm{XOR}}$ computation overhead and $(6N_{\mathrm{CM}} + 4N_{\mathrm{CH}} + 15)|\mathbb{Z}_{p}^{*}| + (6N_{\mathrm{CM}} + N_{\mathrm{CH}} + 3)|\mathrm{T}|$ bits of communication overhead. These results indicate that \textbf{LC2Am} achieves highly lightweight cross-cluster authentication, significantly reducing both computation and communication burdens for EUAVs.

\item \textbf{P2) Low communication overhead for multiple NUAV authentication:}  
When $N$ NUAVs simultaneously initiate the join process, the use of \textbf{MAm} substantially reduces communication overhead. Specifically, the total communication cost is limited to $20|\mathbb{Z}_{p}^{*}| + 4|\mathrm{T}|$ bits, compared to the baseline scheme where authenticating each NUAV individually incurs a cumulative cost of $(7N_{\mathrm{NUAV}} + 3N_{\mathrm{CM}} + 8)|\mathbb{Z}_{p}^{*}| + (2N_{\mathrm{NUAV}} + N_{\mathrm{CM}} + 1)|\mathrm{T}|$ bits. This substantial reduction in communication overhead demonstrates that \textbf{MAm} ensures scalable and efficient authentication for large numbers of NUAVs, making the system well-suited for bandwidth-constrained low-altitude environments.
\end{itemize}

\section{Conclusion}

To ensure the reliability of dynamic Unmanned Aerial Vechiles (UAV) cluster services, specifically in terms of UAV authenticity and the confidentiality of the cluster session key, this paper has proposed the first Lightweight and Privacy-Preserving Cluster Authentication and Session Key Update (LP2-CASKU) scheme. LP2-CASKU has addressed three key challenges: 1) efficient and scalable authentication of new UAVs (NUAVs), 2) privacy-preserving cross-cluster authentication for existing UAVs, and 3) the ensurance of both forward and backward secrecy of cluster session key. The security properties of LP2-CASKU have been rigorously validated through formal analysis, while its computation and communication performance has been assessed via theoretical evaluation and simulation experiments. Results have demonstrated that LP2-CASKU, supported by its proposed message aggregation mechanism, has enabled the concurrent authentication of multiple NUAVs with significantly reduced latency and energy overhead. These findings have confirmed the scheme's practicality and effectiveness in maintaining secure and efficient UAV cluster operations in real-time, bandwidth- and energy-constrained environments characteristic of low-altitude economy networks.

Future work will focus on addressing additional challenges in UAV cluster service provisioning, particularly the design of robust Cluster Head (CH) re-election mechanisms to ensure service continuity in scenarios involving CH failure or mobility-induced disruption.


\begin{thebibliography}{1}

\bibitem{liu2025energy}
Z.~Liu, J.~Zhang, Y.~Zeng, and B.~Ai, ``Energy-efficient multi-agent reinforcement learning for UAV trajectory optimization in cell-free massive MIMO networks,'' \emph{IEEE Transactions on Wireless Communications}, 2025.

\bibitem{jiang2023ntn}
Y.~Jiang, X.~Li, G.~Zhu, H.~Li, J.~Deng, K.~Han, \emph{et al.}, ``6G non-terrestrial networks enabled low-altitude economy: Opportunities and challenges,'' \emph{arXiv preprint arXiv:2311.09047}, 2023.



\bibitem{ref2}
B.~Jiang, J.~Yang, and H.~Song, ``Protecting Privacy From Aerial Photography: State of the Art, Opportunities, and Challenges,'' in \emph{Proc. IEEE INFOCOM Workshops}, 2020, pp. 799--804.

\bibitem{ref3}
P.~I. Radoglou-Grammatikis, P.~G. Sarigiannidis, T.~Lagkas, and I.~D. Moscholios, ``A Compilation of UAV Applications for Precision Agriculture,'' \emph{Comput. Networks}, vol. 172, Art. no. 107148, 2020.

\bibitem{ref4}
A.~Raja, L.~Njilla, and J.~Yuan, ``Adversarial Attacks and Defenses Toward AI-Assisted UAV Infrastructure Inspection,'' \emph{IEEE Internet Things J.}, vol.~9, no.~23, pp. 23379--23389, 2022.

\bibitem{ref5}
Y.~Wan, Y.~Zhong, A.~Ma, and L.~Zhang, ``An Accurate UAV 3-D Path Planning Method for Disaster Emergency Response Based on an Improved Multiobjective Swarm Intelligence Algorithm,'' \emph{IEEE Trans. Cybern.}, vol.~53, no.~4, pp. 2658--2671, 2023.

\bibitem{ref6}
N.~U. I. Hossain, N.~Sakib, and K.~Govindan, ``Assessing the Performance of Unmanned Aerial Vehicle for Logistics and Transportation Leveraging the Bayesian Network Approach,'' \emph{Expert Syst. Appl.}, vol. 209, Art. no. 118301, 2022.

\bibitem{zhang2024moe}
R.~Zhang, H.~Du, Y.~Liu, D.~Niyato, J.~Kang, Z.~Xiong, A.~Jamalipour, and D.~I.~Kim, ``Generative AI agents with large language model for satellite networks via a mixture of experts transmission,'' \emph{IEEE J. Sel. Areas Commun.}, vol.~42, no.~12, pp.~3581--3596, Dec.~2024, doi: 10.1109/JSAC.2024.3459037.


\bibitem{zhang2024generative}
R.~Zhang, H.~Du, D.~Niyato, J.~Kang, Z.~Xiong, A.~Jamalipour, \emph{et al.}, ``Generative AI for space-air-ground integrated networks,'' \emph{IEEE Wireless Communications}, vol.~31, no.~6, pp.~10--20, Dec.~2024.

\bibitem{mao2025covert}
W.~Mao, Y.~Lu, B.~Ai, and T.~Q.~Quek, ``Covert communications in MEC-based networked ISAC systems towards low-altitude economy,'' \emph{arXiv preprint arXiv:2507.18194}, 2025.


\bibitem{ma2025drleisac}
Z.~Ma, R.~Zhang, B.~Ai, Z.~Lian, L.~Zeng, and D.~Niyato, ``Deep reinforcement learning for energy efficiency maximization in RSMA-IRS-assisted ISAC system,'' \emph{IEEE Trans. Veh. Technol.}, early access, 2025, doi: 10.1109/TVT.2025.3580859.


\bibitem{ai2025smart}
B.~Ai, Y.~Lu, Y.~Fang, D.~Niyato, R.~He, W.~Chen, \emph{et al.}, ``6G-enabled smart railways,'' \emph{arXiv preprint arXiv:2505.12946}, 2025.

\bibitem{zhang2023ppo}
R.~Zhang, K.~Xiong, Y.~Lu, P.~Fan, D.~W.~K.~Ng, and K.~B.~Letaief, ``Energy efficiency maximization in RIS-assisted SWIPT networks with RSMA: A PPO-based approach,'' \emph{IEEE J. Sel. Areas Commun.}, vol.~41, no.~5, pp.~1413--1430, May~2023, doi: 10.1109/JSAC.2023.3240707.



\bibitem{ref8}
L.~Zhou, S.~Leng, Q.~Liu, and Q.~Wang, ``Intelligent UAV Swarm Cooperation for Multiple Targets Tracking,'' \emph{IEEE Internet Things J.}, vol.~9, no.~1, pp. 743--754, 2022.

\bibitem{ref9}
P.~Cao, L.~Lei, S.~Cai, G.~Shen, X.~Liu, X.~Wang, L.~Zhang, L.~Zhou, and M.~Guizani, ``computation Intelligence Algorithms for UAV Swarm Networking and Collaboration: A Comprehensive Survey and Future Directions,'' \emph{IEEE Commun. Surv. Tutorials}, vol.~26, no.~4, pp. 2684--2728, 2024.

\bibitem{ref10}
S.~Javed, A.~Hassan, R.~Ahmad, W.~Ahmed, R.~Ahmed, A.~Saadat, and M.~Guizani, ``State-of-the-Art and Future Research Challenges in UAV Swarms,'' \emph{IEEE Internet Things J.}, vol.~11, no.~11, pp. 19023--19045, 2024.

\bibitem{ref11}
L.~Gupta, R.~Jain, and G.~Vaszkun, ``Survey of Important Issues in UAV Communication Networks,'' \emph{IEEE Commun. Surv. Tutorials}, vol.~18, no.~2, pp. 1123--1152, 2016.

\bibitem{ref12}
A.~Perrusquía and W.~Guo, ``Closed-Loop Output Error Approaches for Drone's Physics Informed Trajectory Inference,'' \emph{IEEE Trans. Autom. Control.}, vol.~68, no.~12, pp. 7824--7831, 2023.

\bibitem{cai2025secure}
L. Cai, Y. Zhang, Y. Liu, C. Hu, K. Zhang, B. Yang, Y. Shen, and Z. Yan, ``Secure physical layer communications for low-altitude economy networking: A survey,'' \emph{arXiv preprint arXiv:2504.09153}, 2025.


\bibitem{ref13}
Y.~Tan, J.~Wang, J.~Liu, and N.~Kato, ``Blockchain-Assisted Distributed and Lightweight Authentication Service for Industrial Unmanned Aerial Vehicles,'' \emph{IEEE Internet Things J.}, vol.~9, no.~18, pp. 16928--16940, 2022.

\bibitem{ref14}
C.~Feng, B.~Liu, Z.~Guo, K.~Yu, Z.~Qin, and K.~K. R.~Choo, ``Blockchain-Based Cross-Domain Authentication for Intelligent 5G-Enabled Internet of Drones,'' \emph{IEEE Internet Things J.}, vol.~9, no.~8, pp. 6224--6238, 2022.

\bibitem{ref15}
Z.~Zhang, X.~Li, Y.~Wang, Y.~Miao, X.~Liu, J.~Weng, and R.~H. Deng, ``TAGKA: Threshold Authenticated Group Key Agreement Protocol Against Member Disconnect for UANET,'' \emph{IEEE Trans. Veh. Technol.}, vol.~72, no.~11, pp. 14987--15001, 2023.

\bibitem{ref16}
R.~Karmakar, G.~Kaddoum, and O.~Akhrif, ``A Blockchain-Based Distributed and Intelligent Clustering-Enabled Authentication Protocol for UAV Swarms,'' \emph{IEEE Trans. Mob. Comput.}, vol.~23, no.~5, pp. 6178--6195, 2024.

\bibitem{ref17}
M.~Xie, Z.~Chang, H.~Li, and G.~Min, ``BASUV: A Blockchain-Enabled UAV Authentication Scheme for Internet of Vehicles,'' \emph{IEEE Trans. Inf. Forensics Secur.}, vol.~19, pp. 9055--9069, 2024.

\bibitem{ref18}
I.~Ali, J.~Li, J.~Chen, Y.~Chen, S.~Ullah, and S.~Khan, ``IOOSC-U2G: An Identity-Based Online/Offline Signcryption Scheme for Unmanned Aerial Vehicle to Ground Station Communication,'' \emph{IEEE Internet Things J.}, vol.~11, no.~18, pp. 29941--29955, 2024.

\bibitem{ref19}
W.~Wang, Z.~Han, T.~R. Gadekallu, S.~Raza, J.~Tanveer, and C.~Su, ``Lightweight Blockchain-Enhanced Mutual Authentication Protocol for UAVs,'' \emph{IEEE Internet Things J.}, vol.~11, no.~6, pp. 9547--9557, 2024.

\bibitem{ref20}
M.~Tanveer, H.~Alasmary, N.~Kumar, and A.~Nayak, ``SAAF-IoD: Secure and Anonymous Authentication Framework for the Internet of Drones,'' \emph{IEEE Trans. Veh. Technol.}, vol.~73, no.~1, pp. 232--244, 2024.

\bibitem{ref21}
Omnet++, ``OMNeT++ 6.0.3,'' [Online]. Available: \url{https://omnetpp.org/download-items/omnetpp/omnetpp-603}.

\bibitem{ref22}
K.~S. McCurley, ``The discrete logarithm problem,'' in \emph{Proc. Symp. Appl. Math.}, vol.~42, 1990.

\bibitem{ref23}
D.~R. L. Brown and R.~P. Gallant, ``The Static Diffie-Hellman Problem,'' \emph{IACR Cryptol. ePrint Arch.}, no. 306, 2004.

\bibitem{ref24}
D.~Dolev and A.~C.~C. Yao, ``On the Security of Public Key Protocols,'' \emph{IEEE Trans. Inf. Theory}, vol.~29, no.~2, pp. 198--207, 1983.

\bibitem{ref25}
D.~Stebila, ``An Introduction to Provable Security,'' Lecture Notes, AMSI Winter School on Cryptography, [Online]. Available: \url{https://d1kjwivbowugqa.cloudfront.net/files/teaching/amsi-winter-school/Lecture-2 3-Provable-security.pdf}.

\bibitem{ref26}
V.~Shoup, ``Sequences of Games: A Tool for Taming Complexity in Security Proofs,'' \emph{IACR Cryptol. ePrint Arch.}, no. 332, 2004.

\bibitem{ref27}
J.~De Clercq, ``Single Sign-On Architectures,'' in \emph{Proc. InfraSec}, pp. 40--58, 2002.

\bibitem{ref28}
X.~Xia, S.~M.~M. Fattah, and M.~A. Babar, ``A Survey on UAV-Enabled Edge Computing: Resource Management Perspective,'' \emph{ACM Comput. Surv.}, vol.~56, no.~3, Art. no.~78, 2024.

\bibitem{ref29}
L.~Xu, M.~Chen, M.~Chen, Z.~Yang, C.~Chaccour, W.~Saad, and C.~S. Hong, ``Joint Location, Bandwidth and Power Optimization for THz-Enabled UAV Communications,'' \emph{IEEE Commun. Lett.}, vol.~25, no.~6, pp. 1984--1988, 2021.

\bibitem{ref30}
INET Framework, ``INET 4.5.4 Released,'' [Online]. Available: \url{https://inet.omnetpp.org/2024-10-29-INET-4.5.4-released.html}.

\bibitem{wang2024uavsurvey}
X.~Wang, Z.~Zhao, L.~Yi, Z.~Ning, L.~Guo, F.~R.~Yu, and S.~Guo, ``A survey on security of UAV swarm networks: Attacks and countermeasures,'' \emph{ACM Computing Surveys}, vol.~57, no.~3, pp.~1--37, 2024.

\bibitem{ceviz2024survey}
O.~Ceviz, S.~Sen, and P.~Sadioglu, ``A survey of security in UAVs and FANETs: Issues, threats, analysis of attacks, and solutions,'' \emph{IEEE Communications Surveys \& Tutorials}, early access, 2024.

\bibitem{goldwasser1999}
S.~Goldwasser and M.~Bellare, \emph{Lecture Notes on Cryptography}. Summer course “Cryptography and Computer Security” at MIT, 1999.

\bibitem{albarqi2015pki}
A.~Albarqi, E.~Alzaid, F.~AlGhamdi, S.~Asiri, and J.~Kar, “Public Key Infrastructure: A Survey,” \emph{Journal of Information Security}, vol.~6, pp.~31--37, 2015.

\bibitem{goldreich2001foundations}
O.~Goldreich, \emph{Foundations of Cryptography, Volume 2}, Cambridge University Press, 2004.

\bibitem{katz2007introduction}
J.~Katz and Y.~Lindell, \emph{Introduction to modern cryptography: principles and protocols
}, Chapman \& Hall/CRC, 2007.

\bibitem{vasarhelyi2014flocking}
G.~Vásárhelyi, C.~Virágh, N.~Tarcai, T.~Szőri, G.~Somorjai, T.~Nepusz, and T.~Vicsek,
``Outdoor flocking and formation flight with autonomous aerial robots,''
in \emph{IEEE/RSJ International Conference on Intelligent Robots and Systems (IROS)}, 2014, pp.~3866--3873.

\bibitem{kwa2023density}
H.~L.~Kwa, J.~Philippot, and R.~Bouffanais, 
``Effect of swarm density on collective tracking performance,'' 
\emph{Swarm Intelligence}, vol.~17, no.~3, pp.~253--281, 2023.


\bibitem{bogon2013distuniform}
T.~Bogon, F.~Lorig, and I.~J.~Timm,
``Visualizing the Impact of Probability Distributions on Particle Swarm Optimization,''
in \emph{Advances in Swarm Intelligence (ICSI)}, Lecture Notes in Computer Science, vol.~7928, Springer, 2013.

\bibitem{li2017swarmdensity}
H.~Li, C.~Feng, H.~Ehrhard, Y.~Shen, B.~Cobos, F.~Zhang, K.~Elamvazhuthi, S.~Berman, and A.~L.~Bertozzi,
``Decentralized Stochastic Control of Robotic Swarm Density: Theory, Simulation, and Experiment,''
in \emph{IROS}, 2017.











\end{thebibliography}
\end{document}